\documentclass[letterpaper]{article}

\usepackage{epstopdf}% To incorporate .eps illustrations using PDFLaTeX, etc.
\usepackage[caption=false]{subfig}% Support for small, `sub' figures and tables
\usepackage{mathtools, cases}
\usepackage{blindtext}
\usepackage{varioref}%  smart page, figure, table, and equation referencing
\usepackage{wrapfig}%   wrap figures/tables in text (i.e., Di Vinci style)
\usepackage{threeparttable}% tables with footnotes
\usepackage{dcolumn}%   decimal-aligned tabular math columns
\newcolumntype{d}{D{.}{.}{-1}}
\makeglossary
\usepackage{graphicx}
\usepackage{fancyvrb}
\usepackage{lettrine}
\usepackage{amsmath, amssymb, amsthm}
\usepackage{dcolumn, booktabs, enumitem}

\usepackage{multirow}
\usepackage[table,xcdraw]{xcolor}
\usepackage{algorithm, algcompatible, algpseudocode, algorithmicx}
\usepackage{caption}% http://ctan.org/pkg/caption
\usepackage{xfrac}
\usepackage[export]{adjustbox}
\newcolumntype{L}[1]{>{\raggedright\let\newline\\\arraybackslash\hspace{0pt}}m{#1}}
\newcolumntype{C}[1]{>{\centering\let\newline\\\arraybackslash\hspace{0pt}}m{#1}}
\newcolumntype{R}[1]{>{\raggedleft\let\newline\\\arraybackslash\hspace{0pt}}m{#1}}

\newcommand{\footremember}[2]{%
	\footnote{#2}
	\newcounter{#1}
	\setcounter{#1}{\value{footnote}}%
}

\theoremstyle{plain}% Theorem-like structures provided by amsthm.sty

\theoremstyle{definition}

\theoremstyle{remark}

\makeatletter
\def\hlinewd#1{%
	\noalign{\ifnum0=`}\fi\hrule \@height #1 \futurelet
	\reserved@a\@xhline}
\newcommand{\hthickline}{\hlinewd{1.2pt}}

\algnewcommand\algorithmicforeach{\textbf{for each}}
\algdef{S}[FOR]{ForEach}[1]{\algorithmicforeach\ #1\ \algorithmicdo}
\DeclareMathOperator*{\argmax}{arg\!\max}
\DeclareMathOperator*{\argmin}{arg\!\min}
\renewcommand{\COMMENT}[2][.7\linewidth]{%
	\leavevmode\hfill\makebox[#1][l]{//~#2}}
\algnewcommand\RETURN{\State \textbf{return} }

\newcommand\CONDITION[2]%
{\begin{tabular}[t]{@{}l@{}l@{}}
		#1&#2
	\end{tabular}%
}
\algdef{SE}[WHILE]{While}{EndWhile}[1]%
{\algorithmicwhile\ \CONDITION{#1}{\ \algorithmicdo}}%
{\algorithmicend\ \algorithmicwhile}
\algdef{SE}[FOR]{For}{EndFor}[1]%
{\algorithmicfor\ \CONDITION{#1}{\ \algorithmicdo}}%
{\algorithmicend\ \algorithmicfor}
\algdef{S}[FOR]{ForAll}[1]%
{\algorithmicforall\ \CONDITION{#1}{\ \algorithmicdo}}
\algdef{SE}[REPEAT]{Repeat}{Until}{\algorithmicrepeat}[1]%
{\algorithmicuntil\ \CONDITION{#1}{}}
\algdef{SE}[IF]{If}{EndIf}[1]%
{\algorithmicif\ \CONDITION{#1}{\ \algorithmicthen}}%
{\algorithmicend\ \algorithmicif}%
\algdef{C}[IF]{IF}{ElsIf}[1]%
{\algorithmicelse\ \algorithmicif\ \CONDITION{#1}{\ \algorithmicthen}}

\title{Memetic Algorithm-Based Path Generation for Multiple Dubins Vehicles Performing Remote Tasks}

\author{Doo-Hyun Cho\footremember{alley}{Ph.D. Candidate, Department of Aerospace Engineering, KAIST.} and Han-Lim Choi\footremember{trailer}{Associate Professor, Department of Aerospace Engineering, KAIST.}}
\begin{document}

	\maketitle
	
	\begin{abstract}
		This paper formalizes path planning problem for a group of heterogeneous Dubins vehicles performing tasks in a remote fashion and develops a memetic algorithm-based method to effectively produce the paths. In the setting, the vehicles are initially located at multiple depots in a two-dimensional space and the objective of planning is to minimize a weighted sum of the total tour cost of the group and the largest individual tour cost amongst the vehicles.  While the presented formulation takes the form of a mixed-integer linear program (MILP) for which off-the-shelf solvers are available, the MILP solver easily loses the tractability as the number of tasks and agents grow. Therefore, a memetic algorithm tailored to the presented formulation is proposed. The algorithm features a sophisticated encoding scheme to efficiently. In addition, a path refinement technique that optimizes on the detailed tours with the sequence of visits fixed is proposed to finally obtain further optimized trajectories. Comparative numerical experiments show the validity and efficiency of the proposed methods compared with the previous methods in the literature.
	\end{abstract}

	\section{Introduction}\label{sec:intro}
	Recent decades have observed significant signs of progress in research on automation/autonomy of unmanned vehicles in many different aspects such as mission planning, resource allocation, motion coordination, path planning, low-level control, sensing, and communication~\cite{ceccarelli2007micro, bortoff2000path}.
	In particular, multi-agent aspects of a group of unmanned vehicles have been studied to enhance mission performance and resource utilization~\cite{shima2006multiple}, particularly allowing for heterogeneity in agent capabilities and characteristics ~\cite{choi2010decentralized, ponda2012distributed, sundar2015exact}. One crucial decision to fully take advantage of the extended capability of heterogeneous multiple autonomous vehicles is to design paths/tours for the agents in such a way that optimizes a certain mission performance metric. Limitations in agent motion, payload, and energy~ \cite{cai2014survey, valavanis2014handbook} makes such decision making necessary in practice at the same time incurs complications in the problem with different kinds of constraints.

	%There have been many physical improvements related to robot characteristics compared to the past, such as batteries, computing power, etc., but the insufficient operating limits of unmanned vehicles are still typical constraints that exist when performing missions \cite{cai2014survey, valavanis2014handbook}.
%	In this situation, planning and assigning efficient tours for each vehicle is one of the main research topics.
	The problem of this work's interest is to design paths for (unmanned) vehicles to complete all the tasks in the mission area. While this type of decision making can naturally be approached in the framework of traveling salesman problem (TSP), the particular problem of this paper features a few more complications/sophistications:  (a) tasks can be done in a remote manner --  in other words, a task can be treated as done if an agent just passes nearby; (b) agents are subject to non-holonomic motion constraints; (c) agent capabilities are potentially heterogeneous, and (d) both the total travel cost and load balancing of workload are considered as the performance metric.
	
	The first factor significantly increases the size of decision space -- since an agent can perform a task by passing through many different nearby points and these different options would incur a differing amount of cost, it is not only difficult but also impossible in many cases to define a finite-dimensional decision space. A typical way of handling this indefinite/continuous decision space is to discretize/approximate the problem with the notion of sampling~\cite{obermeyer2012sampling, isaacs2013dubins}. In other words, instead of considering all the nearby points around the task, generating a finite number of sample nodes and focus on the solutions passing through those sample nodes. This allows for the adoption of richer solution schemes developed in the literature, as then the problem belongs to the category of \textit{generalized} TSP. While the handling of arbitrary non-holonomic motion constraints requires calculation of optimal control solutions to determine agent paths, Dubins vehicle model can be adopted if the agent speed can be regarded as constant and the cost metric is the travel time or the path length. A TSP for a Dubins vehicle can be treated as an \textit{asymmetric} TSP, which is still much more complicated than the original TSP.  Multi-agent aspects of the problem necessitate the extension of the aforementioned framework, particularly giving rise to the discussion on the choice of objective function to be optimized. The heterogeneity of agent capabilities incurs additional complexity -- agent to task compatibility often serves as a constraint in the problem and agent speed and maneuverability often affects the cost calculation. While each of the aforementioned complicating aspects has been addressed in the literature as summarized in Section \ref{sec:lit_review}, there has been little work that combined all of these aspects in a systematic fashion. This paper newly suggests to formalize a variant of the traveling salesman problem, termed \textit{generalized, heterogeneous, multi-depot, asymmetric traveling salesman problem (GHMDATSP)} to deal with the aforementioned aspects that are meaningful in practice.
	
	The key contributions of this work are threefold. 	First, this work presents a mixed-integer linear program (MILP) formulation that allows for handling real-world instances of GHMDATSP, full version of which has not been presented in the literature. The formulation  builds upon sampling-based discretization in GTSP and the Dubins vehicle mode, but particularly takes advantage of the notion of \textit{necessarily intersecting neighborhood} (NIN), which was first introduced in the authors' earlier work~\cite{jang2016optimal}, to exclude inefficient tours from the solutino space with consideration of non-holonomic motion constraints.  Second, a memetic algorithm (MA) tailored to the GHMDATSP formulation that computes an optimized tours for a given sent of discretized specification is newly devised; then, a path refinement procedure that further optimizes on the sample nodes is presented to compute a provably better solution than the MA solution. Third, the proposed methods are verified through extensive numerical results in which the performance and computational time are superior to the results from other previously published heuristics.
	
	The rest of the paper is organized as follows.
	Section \ref{sec:lit_review} discusses the relevant literature.
	Section \ref{sec:background} describes in detail the notation and assumptions used in the problem, and a formulation of the GHMDATSP using a mixed integer linear programming with additional valid constraints in a detail.
	The novel memetic algorithm to get a near-optimal solution is shown in Section \ref{sec:ma}.
	Computational results are provided and discussed in Section \ref{sec:num_exp}.
	Finally, Section \ref{sec:conc} gives the conclusion of this study.

	\section{Literature Review}\label{sec:lit_review}
	
	The problem of path planning for multiple unmanned vehicles has been studied in a variety of engineering fields, using numerous kinds of methods.
	It is beyond the scope of this paper to cover all of the existing studies, so we focus on the traveling salesman problem (TSP) and its variants that are directly related to the problem which is suggested in this work, and the approaches to path planning for nonholonomic vehicles.
	
	\subsection{Traveling Salesman Problem and Its Variants}
	
	There have been tremendous research efforts on the TSP related to the path planning \cite{oberlin2010today, sundar2017path}.
	A classic TSP is a problem of finding a visiting order that minimizes the cost of a closed path (or Hamiltonian path) that visits all the list of tasks exactly once given the locations and the distance between each location \cite{dantzig1954solution}.
	It is a problem that does not consider the motion constraints of the vehicle; in other words, it is a suitable model only for a holonomic vehicle that has no restrictions on its movement.
	In most cases, however, the motion constraint of the vehicle necessarily exists because of the inherent kinematic characteristic.
	In particular, a fixed-wing type aerial vehicle only moves forward and can control the change of direction only within a limited range.
	Therefore, when applying the results of the classical TSP to such a nonholonomic vehicle, a large error can occur if the vehicle cannot follow the given path due to its constraints.
	The Dubins TSP, a variant of the TSP, assumes that the motion constraints of the vehicle follow the Dubins model \cite{dubins1957curves} when solving the path planning problem \cite{le2012dubins, isaacs2013dubins, oberlin2009transformation, obermeyer2012sampling, zhang2014memetic}.
	
	In addition, the classic TSP assumes that the vehicle reaches the specified points exactly, but this is rarely required in real-world applications.
	It is natural and desirable to perform tasks slightly off the designated location when the tasks are related to data collection such as monitoring (e.g., crop detection and pollution measurement), event detection (e.g., fire and flood detection), and target tracking (e.g., surveillance and reconnaissance) \cite{cho2016informative, Kim2016cubature, Choi2015informative, yuan2007optimal, liu2013path}.
	A model that can be used appropriately in the above situations is a variant of TSP called generalized TSP (GTSP) or TSP with neighborhoods (TSPN) \cite{gutin2006traveling}.
	GTSP is a generalization of TSP where tasks are substituted with areas or sets of points.
	In the latter case, GTSP can be formulated by creating several sample nodes in the neighborhood area of previously defined tasks.
	The tasks are assumed to be completed when the solution path visits at least one of the task's nodes.
	In addition, if the processing time of the task is short, it can be assumed that the vehicle has performed the task by simply passing near the task, without visiting the sample node.
	It is observed that a TSP with a concept called the intersecting neighborhood (or the necessarily intersecting neighborhood) generated a more effective path than a general TSP when generating a vehicle's path in a situation where remotely executable tasks are densely located \cite{isaacs2013dubins, jang2016optimal}.
	
	In addition to the GTSP, the TSP has been extended in many ways, one of which is the multiple TSP (MTSP) \cite{bellmore1974transformation, laporte1980cutting} where multiple vehicles collaboratively visit the points to fulfill the mission.
	Similar to the TSP, the MTSP has several variants.
	Each vehicle can start its tour with a designated starting point, denoted as the depot.
	If every tour originates from the same point, this problem is called the single depot MTSP.
	Otherwise, the problem is called the multiple depot MTSP (MDMTSP) \cite{gutin2006traveling} if each tour can originate from different points.
	
	Another interesting variant of the MTSP is to construct the objective function using min-max \cite{applegate2002solution, kivelevitch2013market}.
	In other words, the objective function is not to minimize the sum of the costs for all the tours, but to minimize the largest of the tour costs allocated to each vehicle.
	When modeling the objective function in this way, a tour is assigned to each vehicle with almost equal cost.
	The cost is proportional to time in most cases, so the min-max can be interpreted as minimizing the time required to complete the entire mission.
	However, when the objective function is set to the min-max, it is often confirmed that the solver is focused on the maximum vehicle cost and generates an unnecessary tour for the rest of the vehicles.
	
	Furthermore, the MTSP with vehicles with different characteristics is regarded as the heterogeneous MTSP \cite{baldacci2008routing, oberlin2009transformation, sundar2017algorithms}.
	The word heterogeneous can be applied in the sense that the vehicles can differ in the motion constraints from different structures, \textit{structural} heterogeneity, or different task, \textit{functional} heterogeneity, due to the sensor characteristics.
	
	To our knowledge, mathematical formulations and solution methods for the generalized, heterogeneous, multi-depot, asymmetric traveling salesmen problem (GHMDATSP) have never been studied.
	The GHMDATSP can be considered as a generalization of the generalized multi-depot traveling salesmen problem (GMDTSP) \cite{sundar2016generalized} or heterogeneous multi-depot traveling salesmen problem (HMDTSP) \cite{salhi2014multi, sundar2015exact}, which are known to be NP-Hard.

	\subsection{Approaches to the Problems with Motion Constraints}
	
	Most of the studies on the aerial vehicle path planning use the Dubins vehicle model for simplicity, and we also approach the problem with the assumption that vehicles obey the above model.
	The approaches to the DTSP or DTSP with neighborhoods (DTSPN) can be broadly categorized as follows.
	The first class represents decoupling methods that determine the heading angle of each task after determining the visiting order of the given list of tasks.
	The second represents transformation methods in which several heading angles for each task are sampled and then the problem is converted to the asymmetric TSP (ATSP).
	The methods in the third class formulate the problem in the form of mixed integer linear programming, and obtain the optimal solution with exact algorithms such as a branch-and-bound algorithm or a branch-and-cut algorithm.
	The last class represents the methods that exploit evolutionary techniques, such as the genetic algorithm (GA).
	
	The most basic approach to the DTSP is the alternating algorithm (AA) presented in \cite{savla2005point}.
	In the AA, the solution of the classic TSP is fixed as a visiting order.
	Then, the headings of odd-numbered tasks are set to make the straight line segment with each next even-numbered task, and the remaining parts are set to the optimal Dubins paths.
	The upper bound of the solution obtained through the AA is known as $ L_{\textrm{TSP}} \kappa \lceil n/2 \rceil \pi\rho $ where $ L_{\textrm{TSP}} $ is the cost of the optimal solution of the $ TSP $, $ \kappa < 2.658 $, $ n $ is the number of the tasks, and $ \rho $ is the minimum turning radius.
	Similar but slightly more advanced than the AA algorithm, the look-ahead algorithm was presented in \cite{ma2006receding} to determine the heading of each location with using three successive points.
	
	The general steps of the transformation methods are as follows.
	First, the locations and distances are used to generate a complete graph which represents the original problem.
	The graph is then converted into the form of the ATSP.
	After solving the ATSP-formmated graph using a state-of-the-art solver, the output is converted to the original format to obtain the final solution.
	Much research has been done on the transformation methods, and this is one of the major reasons to take advantage of the existing ATSP solver's superior performance.
	In \cite{oberlin2009transformation}, the graph for the heterogenous, multiple depot, and multiple traveling salesman problem (HMDMTSP) was constructed for situations where each of several vehicles has a different turning radius and the heading of each location is given an arbitrary value, and then the graph is converted into the form of the ATSP using the Noon-Bean transformation.
	In \cite{obermeyer2012sampling}, the DTSPN problem was converted into the GTSP with disjoint node sets by generating a number of sample nodes for each task, and finally it was converted into the ATSP.
	The DTSPN problem was converted to the GTSP in \cite{isaacs2013dubins} as it was done in \cite{obermeyer2012sampling}, and the concept called an \textit{intersecting neighborhood} was added to handle the densely located tasks efficiently.
	Similar to \cite{isaacs2013dubins}, the TSPN was handled in \cite{jang2016optimal} by creating a sampling based roadmap, but distances were calculated based on the optimal control approach rather than limiting the dynamics of the vehicle as in the Dubins model.
	They also borrowed the idea called \textit{intersecting neighborhood} from \cite{isaacs2013dubins}, modified it to improve the performance and called as the \textit{necessarily intersecting neighborhood}.
	The GHMDATSP, which we solved in this paper, was handled in \cite{cho2018heterogeneous} using the transformation method and the \textit{necessarily intersecting neighborhood}.
	In each study, the solution of the ATSP was obtained using the LKH algorithm \cite{helsgaun2000effective}.
	
	An exact algorithm has been used to analyze the mathematical characteristics of the problem or to obtain the optimal solution of the instances.
	The mathematical formulation and its branch-and-cut algorithm for the generalized multiple depot multiple traveling salesmen problem was described in \cite{sundar2016generalized}.
	Similar to the above, the HMDMTSP with the Dubins vehicle or the Reeds-Shepp vehicle model was analyzed with the suggested branch-and-cut algorithm in \cite{sundar2017algorithms}.
	Ideally, the exact algorithm would be most advantageous in terms of optimality, but the size of an instance is quite limited due to its inherent limitations.
	
	To overcome the scalability issue, evolutionary computational methods such as the genetic algorithm (GA) and memetic algorithm (MA) have been used in various studies.
	The possibility of task assignment to multiple unmanned aerial vehicles was confirmed through pure GA in \cite{shima2006multiple}.
	In \cite{obermeyer2009path}, the GA was used after the DTSPN with a single vehicle was converted into the GTSP.
	An MA, which combines the pure GA and the local search strategy to find the local optimal heading in each task region, was suggested in \cite{Zhang2014} to solve the DTSPN with multiple Dubins vehicles.
	
%	In this paper, we suggest a mathematical formulation of the GHMDATSP as described in Section \ref{sec:intro}, and then propose the processes of i) obtaining the optimal solution through a branch-and-cut algorithm in Section \ref{sec:algo}, ii) obtaining the near-optimal solution through the novel memetic algorithm in Section \ref{sec:ma}, and iii) refining paths by applying a local optimization to the outputs from i) and ii) in Section \ref{sec:path_refine}.
	
	\section{Statement of the GHMDATSP}\label{sec:background}
	
	\subsection{Dubins Vehicle Model}\label{subsec:dubins}
	
	If a fleet of unmanned aerial vehicles is a fixed-wing type, it can be assumed that each vehicle follows a Dubins vehicle dynamics where the vehicles can only move forward.
	Dubins path refers to the shortest curve in the 2-dimensional Euclidean plane which connects initial and terminal points with given tangents.
	The mathematical model of this system is as follows: $ \dot{x}_k = v_k\cos\theta_k $, $ \dot{y}_k = v_k\sin\theta_k $, and $ \dot{\theta}_k = \frac{\gamma_k}{v_k}u_k $.	
	The subscript $ k \in K=\{1,\cdots,m\} $ denotes the index of a vehicle, and $ m $ is the number of vehicles in a fleet to be coordinated.
	$ x_k $ and $ y_k $ pair is the position of a vehicle $ k $ in a 2-D plane; $ v_k $ is the speed of a vehicle $ k $, and every $ v_k $ is a constant during the entire scenario instance; $ \dot{\theta}_k $, $ \theta_k $ are the angular velocity and heading angle of the vehicle $ k $.
	$ u_k $ is the control input of the vehicle $ k $ to change the heading which varies from -1 to 1.
	Negative and positive values indicate left and right turns, respectively.
	If the vehicle takes coordinate turns in a level flight, a normalization constant of the control input $ \gamma_k $ can be assumed as $ g\sqrt{l^2_{max,k}-1} $ where $ g $ is a gravitational acceleration and $ l_{max,k} $ is a maximum load factor of a vehicle $ k $.
	In this study, the value of $ u_k $ is set as -1, 0, or 1 to make the length of a curve as short as possible.

	\subsection{Notations for the GHMDATSP}\label{subsec:notation}
	
	The goal of the GHMDATSP is to find tours that minimize some global cost when a list of tasks and a heterogeneous fleet of vehicles are given.
	Let $ T = \{1,\cdots,n\} $ be a set of tasks to be visited, and $ D = \{(n+1)^1,\cdots,(n+1)^m,(n+2)^1,\cdots,(n+2)^m\} $ be a set of depots and terminals for each vehicle (the initial and final locations of vehicles); we have a heterogeneous fleet of $ m $ unmanned vehicles initially located at their own depots.
	$ K = \{1,\cdots,m\} $ is used as the notation of a set of vehicles.
	$ (n+1) $ and $ (n+2) $ in $ D $ denote the depot and terminal, respectively.
	Each cardinality of $ T $, $ D $, and $ K $ is denoted as $ |T| = n $, $ |D| = 2m $, and $ |K| = m $.
	
	In a typical GTSP, a set of sample nodes belonging to the same task is referred to as a cluster.
	For each task, depot, and terminal, a cluster, which is a set of sample nodes, are independently created and assigned to each vehicle and task.
	A cluster for vehicle $ k $ and task $ t $ is denoted by $ V^k_t $ where $ k\in K $ and $ t \in T \cup D $, $ V^k $ denotes a union of every task clusters for vehicle $ k $, and $ V_t $ denotes a union of every vehicle clusters for task $ t $.
	A set of every sample node in the instance is denoted by $ V = \bigcup_{k\in K, t\in T \cup D} V^k_t $.
	Given $ V $, we can define a set of directed edges $ E $ between sample nodes belonging to different clusters.
	Note that there is no edge between sample nodes of different vehicles.
	For any given nonempty subset $ S \subset V $ and sample node $ s $, $ t(S) $ and $ t(s) $ are a set of tasks or a task to which the sample nodes in $ S $ or $ s $ belong, respectively.
	We define the $ i^{\textrm{th}} $ sample node in the cluster $ V^k_t $ as $ V^{k}_{t,i} \in V^k_t $.
	In the remainder of this paper, we use $ s\in V^k_t $ instead of $ V^{k}_{t,i} $ for the sake of brevity if possible.
	
	The problem can be formulated on a directed graph $ G=(V,E) $.
	Like the general TSP problem, the defined $ G $ does not include a self-loop.
	For each directed edge $ (s, s') \in E $ when the sample nodes belong to the vehicle $ k $, $ c_{s,s'} $ is the cost for vehicle $ k $ of traversing from $ s $ to $ s' $.
	The definition of cost depends on the purpose of the mission, but can generally be defined as the time or distance traveled by the vehicle along the path or the amount of fuel used.
	The cost can be calculated with the model in Section \ref{subsec:dubins}.	
	
	\subsection{Necessarily Intersecting Neighborhoods}\label{subsec:NIN}

	\begin{figure}
		\centering
		\begin{minipage}{.8\textwidth}\centering
			\includegraphics[width=\textwidth]{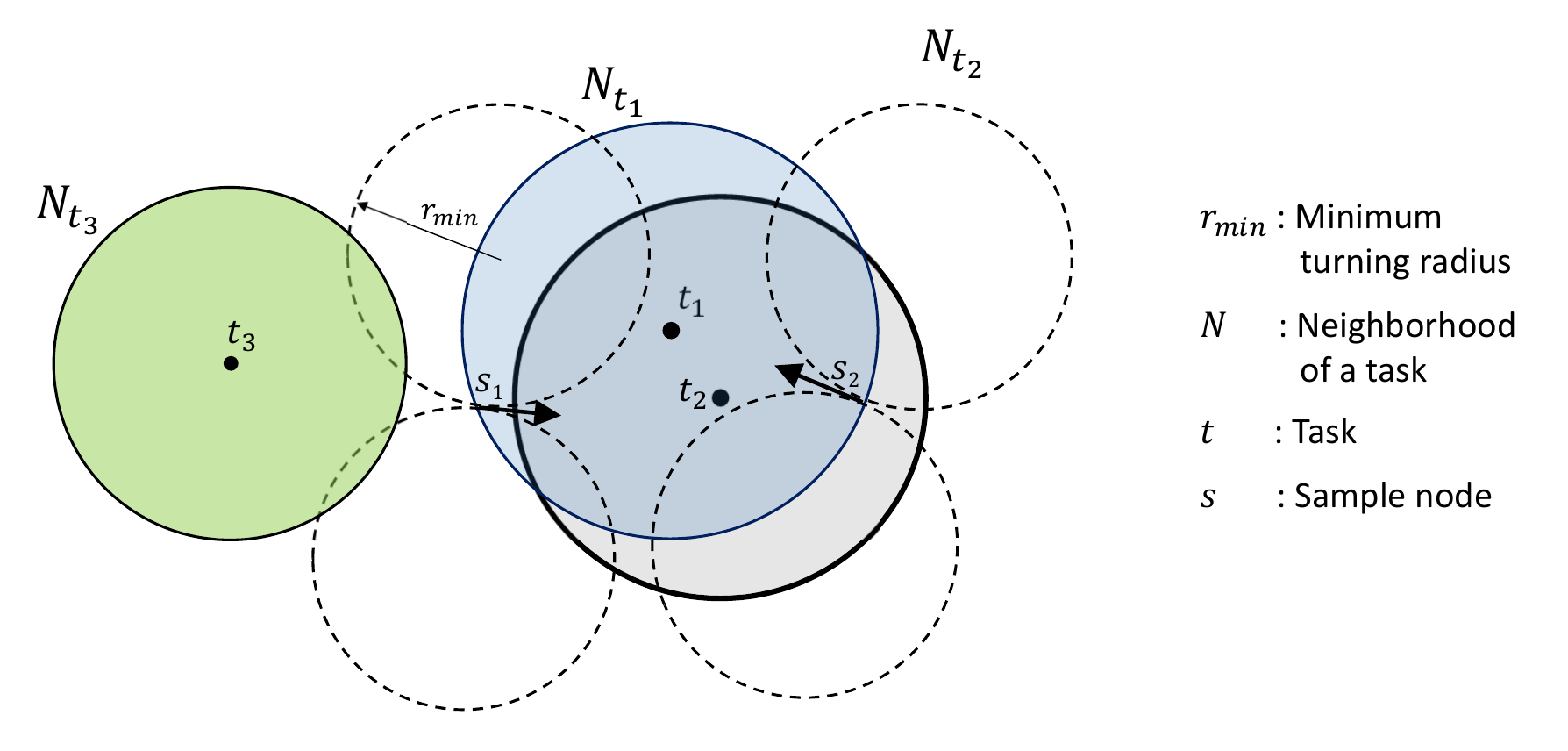}
			\caption{An example of a vehicle passing neighborhoods necessarily when entering the sample nodes $ s_1 $ or $ s_2 $.}
			\label{fig:nin}
		\end{minipage}
	\end{figure}
	
	We use the concept called necessarily intersecting neighborhoods (NIN) proposed in \cite{jang2016optimal}, which is an extension of the Intersecting Regions Algorithm \cite{isaacs2013dubins}.
	An instance is illustrated in Figure \ref{fig:nin}  for the explanation of the NIN.
	This instance has a total of three tasks, and sample nodes $ s_1 $ and $ s_2 $ belong to task $ t_1 $; sample nodes for $ t_2 $ and $ t_3 $ are omitted for simplicity.
	Node $ s_1 $ and $ s_2 $ are located at the boundary of the task's neighborhood region $ N_{t_1} $, and the direction of each node is set toward $ t_1 $.
	The vehicle has motion constraints based on the Dubins model, assuming that the minimum turning radius is $ r_{min} $.
	If the vehicle visits the sample node $ s_1 $, it necessarily passes through the region of tasks $ t_2 $ and $ t_3 $ due to its motion constraints.
	Similarly, the vehicle necessarily passes $ t_2 $ if it is set to visit $ s_2 $.
	
	The following is how to verify whether a sample node $ s $ passes through an arbitrary task $ t $.
	Draw two circles with a radium $ r_{min} $ that tangent to $ s $, and then check that a region of $ t $ intersects both circles simultaneously.
	If the above is satisfied, the vehicle necessarily passes $ t $ when visiting $ s $, or in other words, the neighborhood of $ t $ can be expressed as a neighborhood which necessarily intersects $ s $.
	In this paper, a set $ S^{\textrm{NIN}}_t $ and $ T^{\textrm{NIN}}_s $ are created for each task $ t $ and sample node $ s $ respectively.
	In the case of $ S^{\textrm{NIN}}_t $, every sample node except the nodes which belong to the task cluster $ t $ is included in $ S^{\textrm{NIN}}_t $ if $ s $ necessarily intersects $ t $.
	In Figure \ref{fig:nin}, $ S^{\textrm{NIN}}_{t_2} $ is $ \{s_1, s_2\} $, and $ S^{\textrm{NIN}}_{t_3} $ is $ \{s_1\} $.
	Similarly, in the case of $ T^{\textrm{NIN}}_s $, every task $ t $ is included in $ T^{\textrm{NIN}}_s $ except the task which already has the sample node $ s $ if $ t $ is necessarily intersected by $ s $.
	Therefore, $ T^{\textrm{NIN}}_{s_1} $ is $ \{t_2, t_3\} $, and $ T^{\textrm{NIN}}_{s_2} $ is $ \{t_2\} $ in Figure \ref{fig:nin}.

	\subsection{Problem Formulation}\label{subsec:prob_form}
	
	For the formulation, we define the binary decision variables $ \textbf{x} $, $ \textbf{y} $, and $ \textbf{y}^{\textrm{NIN}} $.
	An element of $ \textbf{x} $, $ x_{s,s'} $ is defined for each edge $ (s,s') \in E $, whose value equals 1 if it is chosen as an element of a tour solution and 0 otherwise.
	Using the variable $ \textbf{x} $, a sum of edge cost for vehicle $ k $, $ \textrm{Cost}_k $, is $ \sum_{s,s'\in V^k}c_{s,s'}\cdot x_{s,s'} $.
	An element of $ \textbf{y} $, $ y_s $ is defined whose value is equal to 1 if the node $ s $ is visited by a vehicle.
	Similarly, an element of $ \textbf{y}^{\textrm{NIN}} $, $ y^{\textrm{NIN}}_{t,s} $ is defined to be 1 if sample node $ s $ necessarily intersects task $ t $.
	
	The problem is formulated as follows:
	\begin{alignat}{3}
		\MoveEqLeft[3] \text{Minimize: }\notag\\
		& \qquad \qquad \label{eq:obj} \alpha \dfrac{\left(\sum_{k \in K} \textrm{Cost}_{k}\right)}{m}+(1-\alpha)\max_{k\in K}\textrm{Cost}_{k} \\[1ex]
		\MoveEqLeft[3] \text{subject to}\notag\\
		& \label{eq:vtassign1} y^{\textrm{NIN}}_{t,s} = y_s & \forall t \in T, s \in \textrm{NIN}_{t} \\
		& \label{eq:vtassign2} \left(\sum_{s \in V_t} y_s\right) \vee \left(\bigvee_{s\in \textrm{NIN}_t} y^{\textrm{NIN}}_{t,s}\right) = 1  & \forall t \in T \\
		& \label{eq:vtassign3} \sum_{s \in V^k_t} y_s = 1    &	\forall k \in K, t \in D \\
		& \label{eq:deg} \sum_{s' \in V^k_{T\cup D\setminus \{t(s)\} }} x_{s',s} + x_{s,s'} = 2y_s & \forall k \in K, s \in V^k \\
		& \label{eq:subelim} \sum_{s' \in V\setminus S} x_{s',s} + x_{s,s'} \geq 2y_s \qquad & \forall S\subseteq \bigcup_{t\in T}V_t, s\in S
%		& \label{eq:add} \sum_{\substack{s_1 \in V^k_{n+1}, \\ s_2 \in V^k_{n+2}}} x_{s_1,s_2} = 1 \rightarrow y_s = 0 & \forall k \in K, t \in T, s \in V^k_t
	\end{alignat}
	
	The objective function in Eq. \eqref{eq:obj} is defined to minimize the linear combination of two terms: \textbf{a)} \textit{the mean cost of vehicles} and \textbf{b)} \textit{the maximum cost of a vehicle from a fleet}.
	The $ \alpha $ in Eq.\eqref{eq:obj} is a coefficient that determines which of the above two terms to focus more on to optimize the problem.
	Instead of using the total cost sum at the first term, the mean value is used to normalize the size with respect to the second term.
	Constraints \eqref{eq:vtassign1} bind all the tasks associated with sample node $ s $.
	If the sample node $ s $ is visited so the value equals 1, then all of the necessarily intersecting tasks related to $ s $ are assumed to be visited.
	Constraints \eqref{eq:vtassign2} ensure that each target is visited by some vehicle directly visiting the generated sample node or indirectly visiting with the concept of NIN.
	Constraints \eqref{eq:vtassign3} imply that one of the sample nodes in the depot or the terminal cluster for each vehicle has to be chosen.
	Constraints \eqref{eq:deg} is called the degree constraints.
	If a sample node $ s $ of a vehicle $ k $ is chosen to be visited and the value of $ y_s $ equals 1, the in-degree and out-degree of the node is 1 respectively.
	In other words, one of the edges towards the node $ s $ and one of the outward edges from the node $ s $ should be selected.
	Constraints in \eqref{eq:subelim} prevents the generation of subtours of any subset of tasks for each vehicle.
%	Constraints in \eqref{eq:add} reduce a huge search space during the solving phase especially when the objective function is only to minimize the global sum of the costs.
%	For a vehicle which has not been assigned a task, the vehicle is assigned a tour that connects only the depot and terminal clusters.
%	Since no task sample node associated with the vehicle is visited, the value of $ y_s $ equals zero.
%	As described above, $ n + 1 $ and $ n + 2 $ in eq. \eqref{eq:add} denote depot and terminal, respectively.

	\section{Memetic Algorithm based Path Generation}\label{sec:ma}
	
	\begin{figure}[]
		\centering
		\includegraphics[width=.9\linewidth]{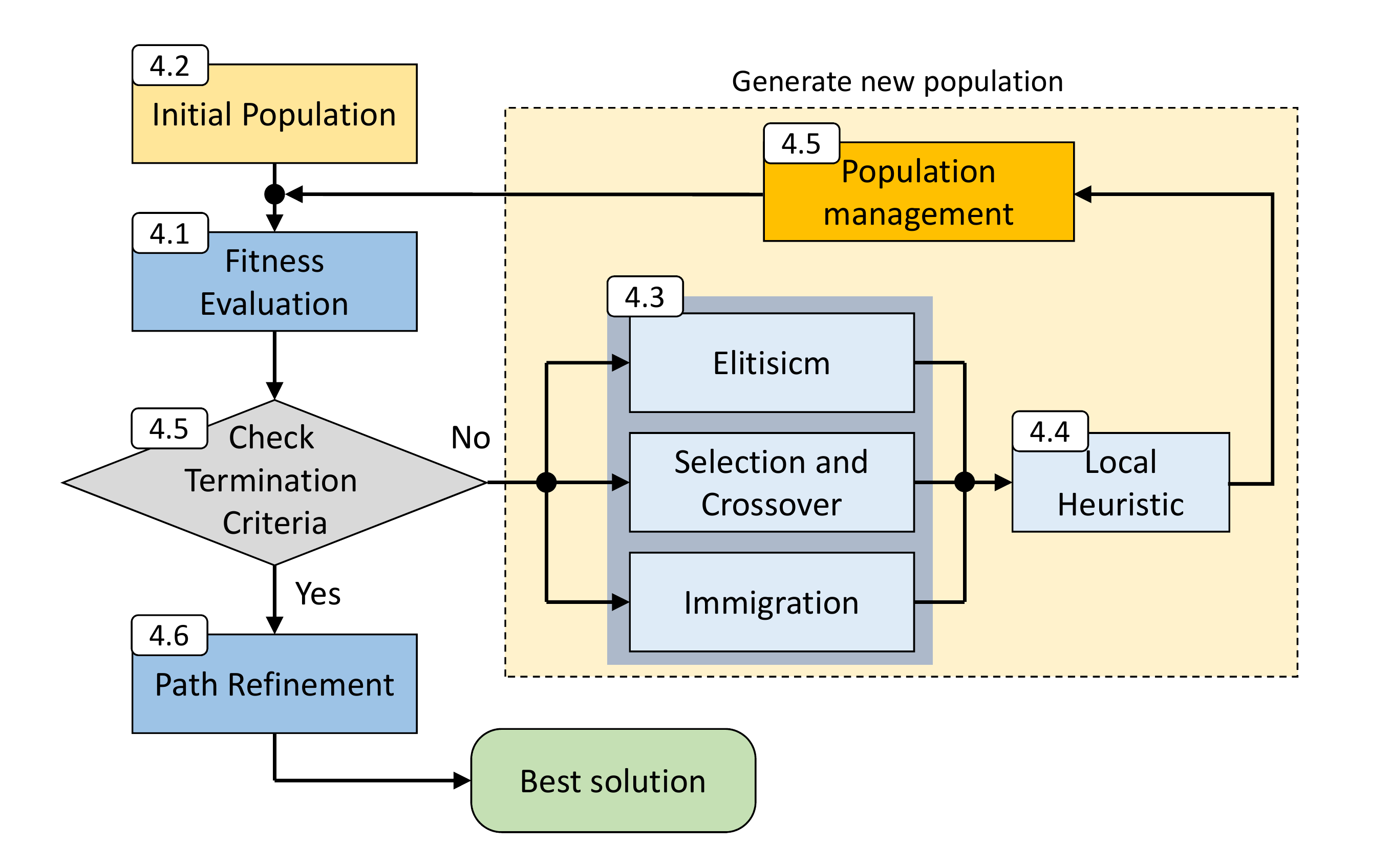}
		\caption{A schematic of the memetic algorithm based path generation procedure.}	
		\label{fig:ga_diagram}
	\end{figure}
	
%	As mentioned above, in addition to the exact algorithm mentioned in Section \ref{sec:algo}, we used the memetic algorithm as an alternative to solve the instances of the GHMDATSP.
	The classical genetic algorithm (GA) is a process that repeatedly evolves a population of chromosomes (or solutions) through operators such as selection, crossover, and mutation, and finally obtains a high-quality solution at the end of the iteration (or generation).
	While borrowing the methodology used in the classical GA, a procedure we propose in this paper consists of modified GA operators, local heuristic methods for chromosomes, and the path refinement process.
	The schematic of the path generation procedure is shown in Figure \ref{fig:ga_diagram}, including related subsection index for each block.
	The following describes each operator that makes up the procedure.

	\subsection{Encoding and Decoding}\label{subsec:encode_decode}
	
	\begin{figure}[]
		\centering
		\includegraphics[width=.7\linewidth]{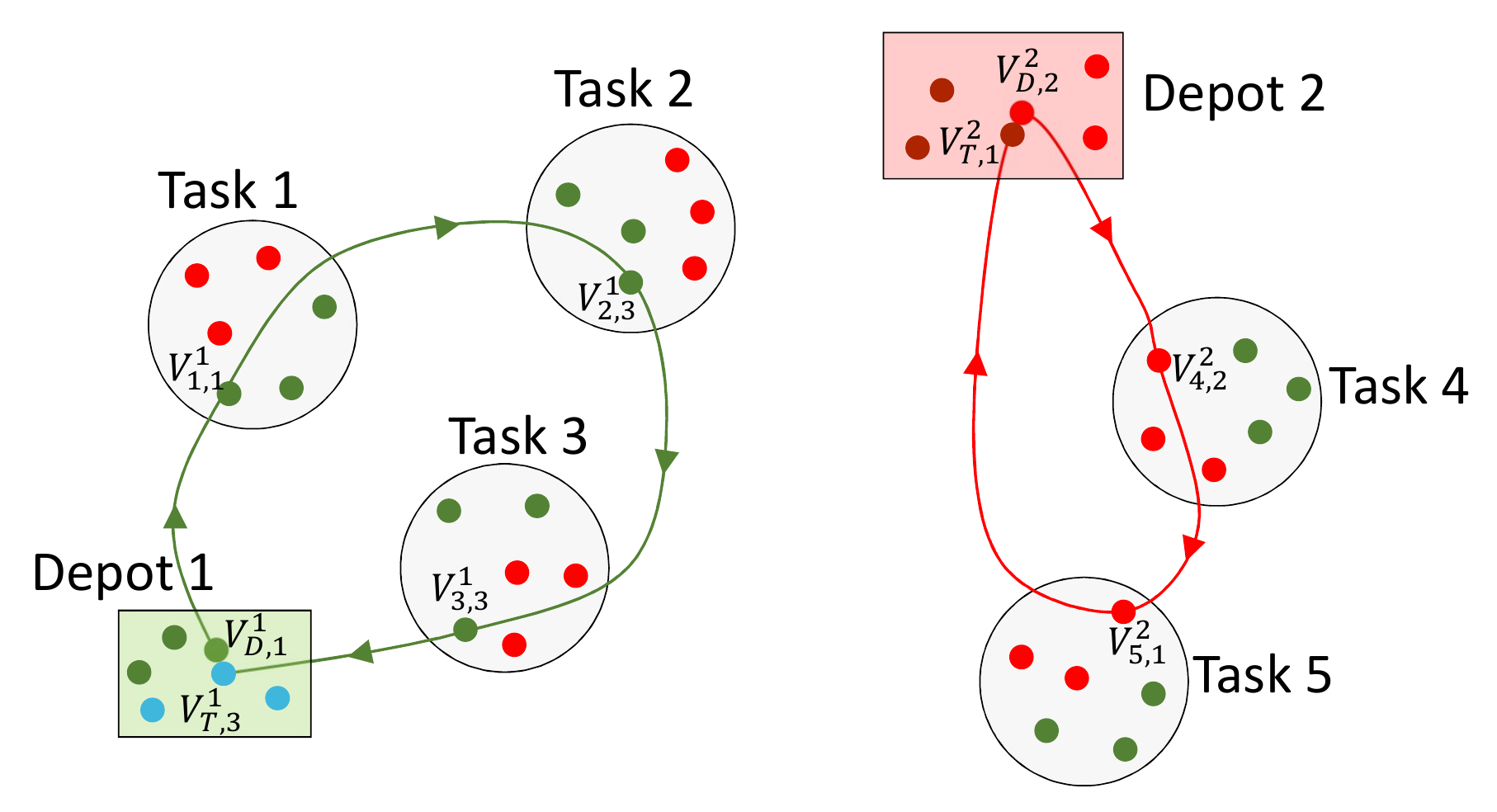}
		\caption{A simple instance of the GHMDATSP.}	
		\label{fig:ga_instance}
	\end{figure}
	
	\begin{figure*}[]
		\centering
		\captionsetup{justification=centering}
		\begin{minipage}{\linewidth}\centering
			\subfloat[An instance with a given chromosome.]{\label{fig:decode_nin_1}\includegraphics[frame,page=1,width=0.42\linewidth]{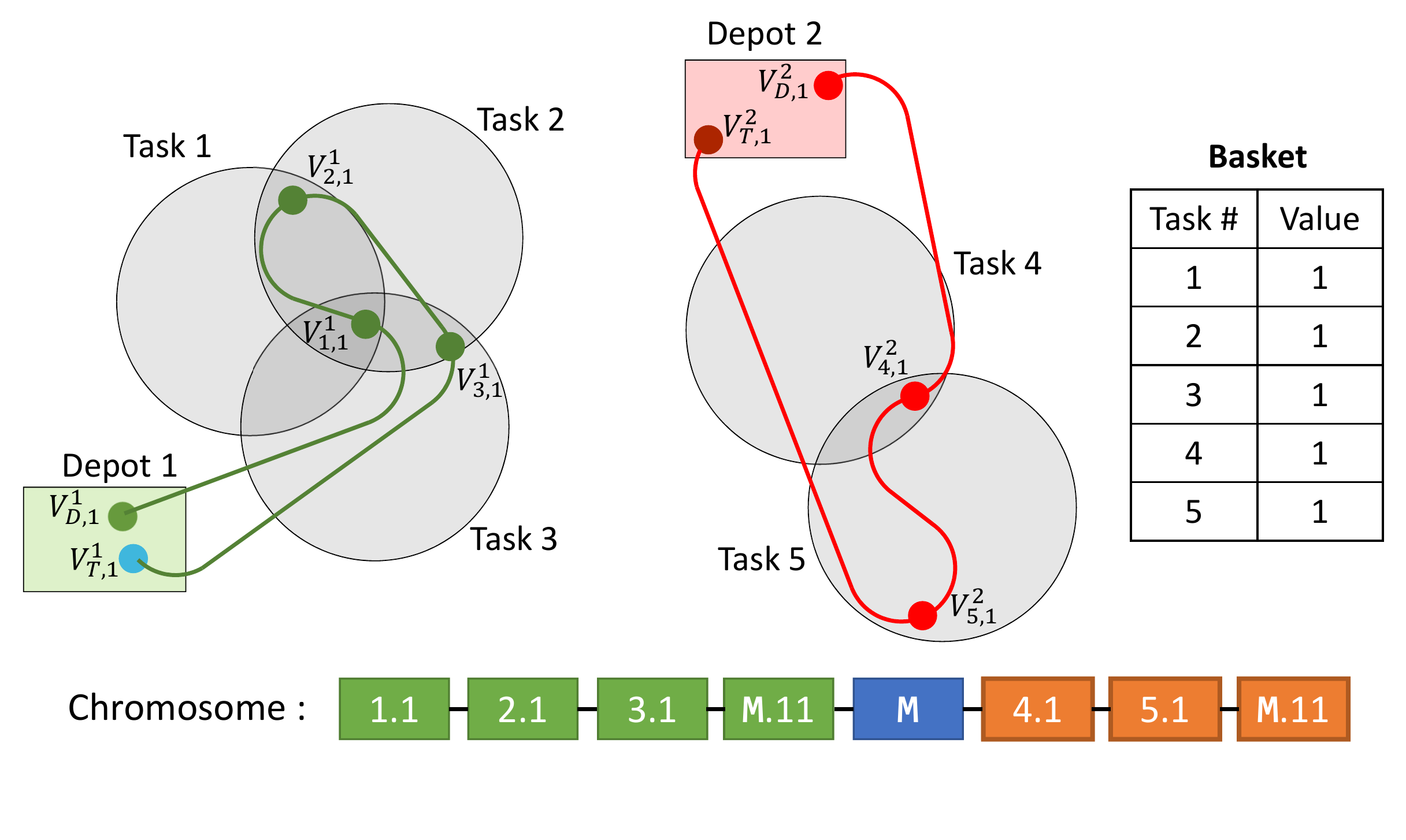}}
			\subfloat[Updating with gene $ 1.1 $.]{\label{fig:decode_nin_2}\includegraphics[frame,page=2,width=0.42\linewidth]{decode_nin}}
		\end{minipage}
		\begin{minipage}{\linewidth}\centering
			\subfloat[Updating with gene $ 2.1 $.]{\label{fig:decode_nin_3}\includegraphics[frame,page=3,width=0.42\linewidth]{decode_nin}}
			\subfloat[Updating with gene $ 3.1 $.]{\label{fig:decode_nin_4}\includegraphics[frame,page=4,width=0.42\linewidth]{decode_nin}}
		\end{minipage}
		\begin{minipage}{\linewidth}\centering
			\subfloat[Updating with gene $ 4.1 $.]{\label{fig:decode_nin_5}\includegraphics[frame,page=5,width=0.42\linewidth]{decode_nin}}
			\subfloat[Updating with gene $ 5.1 $.]{\label{fig:decode_nin_6}\includegraphics[frame,page=6,width=0.42\linewidth]{decode_nin}}
		\end{minipage}
		\begin{minipage}{\linewidth}\centering
			\subfloat[Deleting $ V^1_{2,1} $.]{\label{fig:decode_nin_7}\includegraphics[frame,page=7,width=0.42\linewidth]{decode_nin}}
			\subfloat[Deleting $ V^2_{5,1} $.]{\label{fig:decode_nin_8}\includegraphics[frame,page=8,width=0.42\linewidth]{decode_nin}}
		\end{minipage}
		\subfloat[Deleting $ V^1_{3,1} $.]{\label{fig:decode_nin_9}\includegraphics[frame,page=9,width=0.42\linewidth]{decode_nin}}
		\caption{An example of \textit{decode-NIN} operator.}		
		\label{fig:decode_nin}
	\end{figure*}
	
	Chromosome encoding is the most important part of the memetic algorithm.
	The genes that make up the chromosome are real numbers made up of the sum of integer and fractional parts.
	The integer part stores the task cluster index, and the fractional part stores information about the sample node index.
	We first discuss the general chromosome encoding and decoding method and then describe the decoding method which includes the NIN, which is discussed in \ref{subsec:NIN}.
	
	For example, assume that there are two vehicles, five task clusters, and three sample nodes in each task cluster for each vehicle, as shown in Figure \ref{fig:ga_instance}.
	The following chromosome
	\[ M.13 - 1.1 - 2.3 - 3.3 - M - M.21 - 4.2 - 5.1 \]
	is decoded for each vehicle as follows:
	\begin{itemize}
		\item Vehicle 1: (D,1) - (1,1) - (2,3) - (3,3) - (T,3)
		\item Vehicle 2: (D,2) - (4,2) - (5,1) - (T,1)
	\end{itemize}
	The first element of each tuple is the task, and the second is the index of the sample node.
	D is the depot, and T is the terminal cluster.
	The encoding operator of this study is different from the other studies dealing with GTSP in that it adopts the concept of delimiter to represent multiple vehicles on the chromosome.
	For a multi-vehicle GTSP instance where the number of vehicles is $ k $, the chromosome has $ 2k-1 $ delimiters, and the gene corresponding to the delimiter is assigned $ M $ to the integer part.
	Here $ M $ is a moderately large integer satisfying $ M > n $.
	Thus, the length of the chromosome of a multi-vehicle GTSP with $ k $ vehicles and $ n $ tasks is $ (n + 2k-1) $.
	In the case of the gene corresponding to the delimiter, the roles of the delimiters located at odd-numbered and even-numbered positions are different.
	In the case of the odd-numbered delimiter, it indicates the depot and terminal cluster of the vehicle.
	It also contains information about the sample node of the depot and terminal cluster in the fractional part.
	A delimiter located at an even-numbered position acts as a divider to distinguish the path of the next vehicle from the path of the previous vehicle.
	In the above chromosome, the first and sixth genes contain the depot sample node and terminal sample node information of vehicles 1 and 2, respectively, and the fifth gene divides the paths of vehicles 1 and 2.
	When the order of the delimiters in the chromosome is changed through the operators described in \ref{subsec:imp_heuristics}, the sample node visit information may be shifted to the delimiters at even-numbered positions.
	In this case, the fractional part values of the delimiters that need to be moved are appropriately shifted to another delimiter so that the fractional part exists only in the odd-numbered delimiter.
	A detailed explanation is given in \ref{subsubsec:global_2opt}.
	The decode operator takes a chromosome as an input and outputs a multi-list variable called \textbf{\textit{tour}}=\{\textit{tour}$_1$, $ \cdots $, \textit{tour}$_m$\} consisting of m lists containing tour information for each vehicle.
	
	In order to apply the NIN discussed in \ref{subsec:NIN} to the GA, an additional decoding process is needed in addition to the above, and the decode operator including the additional process is \textit{decode-NIN}.
	The basic decoding scheme of decode-NIN is the same as that described above.
	In addition, the purpose of this decoding operator is to create a reduced \textbf{\textit{tour}} that removes redundant sample nodes from the \textbf{\textit{tour}} through the NIN, so that only a minimum sample node can be visited to traverse all task areas.
		
	For the convenience of explanation, we explain the decode-NIN using Figure \ref{fig:decode_nin}.
	There are 5 tasks and 2 vehicles in a given instance, and there is one sample node for each task cluster.
	Using the basic decode operator, the given chromosome $ 1.1-2.1-3.1-M.11-M-4.1-5.1-M.11 $ has tour information for vehicle 1 to visit sample nodes in tasks 1, 2, and 3, and for vehicle 2 to visit sample nodes in tasks 4 and 5.
	Suppose that each task cluster has one basket.
	Here, basket contains information about how many times the task is visited from all sample nodes of the chromosome.
	Basically, each task has a default value of 1, because each task is unconditionally visited once by the sample node belonging to its cluster (Figure \ref{fig:decode_nin_1}).
	After that, we check the additional visit based on the information of the sample node in the chromosome.
	Since $ V^1_{1,1} $ goes through tasks 2 and 3 simultaneously, the values of baskets 2 and 3 are increased by one(Figure \ref{fig:decode_nin_2}) and similarly the value of basket 1 is increased because $ V^1_{2,1} $ passes task 1 at the same time (Figure \ref{fig:decode_nin_3}).
	A sample node that does not pass other tasks simultaneously, such as $ V^2_{5,1} $, does not affect the basket value of other tasks.
	By checking all the sample nodes in the chromosome in this way and adding the values to the basket, the results shown in Figure \ref{fig:decode_nin_6} can be obtained.
	When the NIN check for all sample nodes in the chromosome is completed, the sample nodes of the tour are deleted as much as possible until all the values of the basket are kept at 1 or more.
	The sample nodes are deleted in the order of the largest value of the basket.
	If there are several tasks with the same basket value, a sample node to be erased is determined by the cardinality of $ T^{\textrm{NIN}}_s $ with the smallest value.
	This is because a smaller $ T^{\textrm{NIN}}_s $ means that the corresponding sample node goes through fewer tasks at the same time and it is better to erase it.
	In Figure \ref{fig:decode_nin_6}, because the basket value of task 2 is the largest, delete the sample node $ V^1_{2,1} $ of task 2 from the tour, and decrease the basket value of tasks 1 and 2, which are the tasks passed through $ V^1_{2,1} $, by one (Figure \ref{fig:decode_nin_7}).
	Since the task with the largest basket value is 3 and 5, it is needed to compare the size of each $ T^{\textrm{NIN}}_s $.
	$ T^{\textrm{NIN}}_{V^1_{3,1}} = {2} $, and $ T^{\textrm{NIN}}_{V^2_{5,1}} = \emptyset $, so a sample node with a smaller $ T^{\textrm{NIN}}_s $ size $ V^2_{5,1} $ is chosen to be deleted and reducing the basket value by one (Figure \ref{fig:decode_nin_8}).
	Finally, we delete the sample node $ V^1_{3,1} $ of task 3 and reduce the basket value of tasks 2 and 3, which are the tasks passed through $ V^1_{3,1} $, by 1 (Figure \ref{fig:decode_nin_9}).
	The tour of \ref{fig:decode_nin_9} becomes the final tour of the given chromosome because if we reduce the sample node once again, there is a task whose basket value becomes zero.
	The Algorithm \ref{alg:decode-nin} describes a pseudo algorithm of the decode-NIN operator.
	
	\begin{algorithm}[t]
		\caption{Decode-NIN}
		\begin{algorithmic}[1]
			\Procedure{Decode-NIN (chromosome, $ \textrm{NIN} $)}{}
			\State \textit{\textbf{tour}} $ \leftarrow $ decode(\textit{chromosome})
			\For{$ t := 1 $ to $ n $}
			\State \textit{basket}$ [t] \leftarrow 1 $
			\EndFor
			\ForEach {sample node $ s $ in \textit{\textbf{tour}}}
			\ForEach {$ t $ in $ T^{\textrm{NIN}}_s $}
			\State \textit{basket}[$t$]++
			\EndFor
			\EndFor
			\While {every value in \textit{basket} is bigger than zero}
			\State $ \textbf{t}^{\textrm{del}} \leftarrow \argmax $(\textit{basket})
			\If {$ \left| \textbf{t}^{\textrm{del}} \right| = 1$}
			\State $ t^{\textrm{del}} = \textbf{t}^{\textrm{del}} $
			\Else
			\State $ t^{\textrm{del}} \leftarrow \argmin_t\left| T^{\textrm{NIN}}_{V^k_{t,i}} \right|$  for t in $ \textbf{t}^{\textrm{del}} $
			\State \COMMENT {$ k $: vehicle visiting $ t $}
			\State \COMMENT {$ i $: sample node index belonging to $ t $}
			\EndIf
			\State {$ s \leftarrow $ sample node belonging to task $ t^{\textrm{del}} $}
			\ForEach {task $ t $ in $ T^{\textrm{NIN}}_s $}
			\State basket[$t$] - -
			\EndFor
			\If {any element of \textit{basket} is zero}
			\State break \textbf{while} loop
			\EndIf
			\State $ k \leftarrow $ vehicle visiting $ s $
			\State delete $ s $ in \textit{tour}$ _k $
			\EndWhile
			\RETURN \textit{\textbf{tour}}
			\EndProcedure
		\end{algorithmic}
		\label{alg:decode-nin}
	\end{algorithm}

	\subsection{Initialization}\label{subsec:initialization}
	
	Two different ways are used to construct the initial population with N chromosomes.
	The first is to generate a sequence of which vehicle to visit which task according to the encoding rules in \ref{subsec:encode_decode}.
	The second is to assign the tasks corresponding to each Voronoi cell of a vehicle after the region of interest $ Q \in \mathbb{R}^2 $ is partitioned into Voronoi cells using the depot locations.
	The Voronoi diagram $ \mathcal{V}(P,Q) $ consists of the $ m $ disjoint Voronoi cells $ \mathcal{V}_i(P,Q) $ generated by the set of points $ P $, in other words, $ \mathcal{V}(P,Q) = \cup_{i=1}^{m}\mathcal{V}_i(P,Q) $.
	An arbitrary task point $ q \in Q $ belongs to $ \mathcal{V}_i(P,Q) $ if $ ||q-p_i|| \leq ||q-p_k|| $ $ \forall k \in {1,\cdots,m} $.
	After the tasks are assigned to the vehicle, the visiting order is sorted by applying the LKH heuristic \cite{helsgaun2015solving}.
	The points in the cell is are given as the input of the ETSP, then its solution is used as the sequence of the initial guess.
	The sample nodes of each task are chosen randomly.
	After constructing the initial population, Level-I improvement in \ref{subsec:imp_heuristics} is applied to complete the initialization.
	
	\subsection{Operators for the MA}\label{subsec:operators}
	
	\subsubsection{Reproduction}
	
	In this study, we apply an \textit{elitist strategy} to the reproduction operator.
	This prevents inefficient behavior so the algorithm finds high quality solutions repeatedly.
	This is done by moving the fittest chromosome group from the previous generation to the next generation.
	Through this strategy, the quality of the overall population is promoted as the generation continues, which ensures that a high quality chromosome is selected in the subsequent selection operator.
	
	\subsubsection{Selection}
	
	The selection operator selects two parent chromosomes as the inputs of the crossover operator.
	Among the various selection operators used in the GA, we used the roulette wheel selection (fitness proportionate selection), which is known as the most representative method.
	The roulette wheel selection is a method that evaluates the cost of each solution and then selects the chromosome by adjusting the fitness of the best solution to be $ \kappa $ times the fitness of the worst solution.
	Here $ \kappa $ is called the selection pressure.
	The higher the selection pressure, the faster convergence but the higher the likelihood of premature convergence.
	If the selection pressure is too low, on the other hand, there is a tendency for the average cost of the population not to improve rapidly.
	$ \kappa $ is an adjustable parameter and is set to 4 in this study.
	The fitness $ f_i $ of each chromosome can be obtained using the following equation: $ f_i = c_w - c_i + \frac{c_w - c_b}{\kappa-1} $ where $ c_w $ and $ c_b $ are the worst and best cost in the pool, and $ c_i $ is the cost of the current chromosome.
	The probability $ p_i $ that each chromosome is selected can be expressed as follows: $ p_i = \frac{f_i}{\sum_{j=1}^{N}f_j} $.
	
	\subsubsection{Crossover}
	
	After selecting two different parent chromosomes using the above selection operator, we create a child chromosome through a parameterized uniform crossover.
	The child chromosome is generated by receiving 60\% of genes from parent 1 and 40\% of genes from parent 2, but excluding the gene that overlaps with the gene of parent 1 based on task cluster (integer part).
	If there is a blank space in the child chromosome because of the redundancy, the sequence of task clusters is randomly arranged, and the sample nodes follow the information of parent 1.

	\subsubsection{Immigration}\label{subsubsec:immigration}
	
	After the next generation of populations is populated with reproduction and crossover operators, the remaining number of chromosomes are generated according to the method described in \ref{subsec:initialization}.
	In this study, the role of mutation operator, which is commonly used in the GA, is replaced with an immigration operator to guarantee a diversity of the solution.

	\subsection{Improvement Heuristics}\label{subsec:imp_heuristics}
	
	Local improvement heuristics play a major role in enhancing the quality of the solution in the general GA.
	In the proposed algorithm, the improvement heuristic is applied to the initial population or to the chromosome newly generated in the crossover or immigration phase.
	Improvement heuristics consists of 2-opt and swap methods, each of which is summarized as follows.
	The 2-opt method consists of a global 2-opt, which applies 2-opt for the entire chromosome, and a local 2-opt applying 2-opt for each tour assigned to the vehicle.
	The swap consists of a task swap that changes the order of two genes across the chromosome and a sample swap that optimizes the sample node information for each gene.
	The main idea of 2-opt, one of the simplest local search algorithms to solve the TSP problem, is that the tour can be re-arranged by tweaking a part of the tour, thereby improving costs.
	For the ETSP (the cost from node $ a $ to node $ b $ is the same as the cost from $ b $ to $ a $), the cost of the entire tour can be updated using only the cost variation between new edges caused by exchanging selected edges and destinations of two different edges.
	However, for the asymmetric TSP, the cost changes according to the direction of edges.
	Therefore, the cost of all the changed paths must be calculated to update the cost of the entire tour.
	To cope with this problem, \cite{freisleben1996genetic} used 3-opt instead of 2-opt to preserve the direction of the path.
	However, we used 2-opt since the number of task clusters to visit is not very large in the instance handled in this study.
	In addition, the algorithm is constructed to guarantee the quality of the chromosome through the swap operators.
	The 2-opt and swap operations are not integrated as used in \cite{renaud1998efficient}, but are made to work as separate operators.
	As in \cite{snyder2006random}, level-I improvement is applied when the cost of the chromosome does not belong to the upper rank, and level-II improvement is applied when it belongs to it.
	We applied the global 2-opt, local 2-opt, and sample swap operators to the chromosome once through Level-I improvement, and we applied the task swap operator 5 times.
	Level-II improvement is applied to the chromosome until the global 2-opt, local 2-opt, task swap operator fail to improve the cost ten times in succession, and the sample swap operator is applied three times.
	The algorithm is designed to make the calculation more efficient by applying the improvement operator intensively to the high quality chromosome.
	
	\subsubsection{Global 2-opt}\label{subsubsec:global_2opt}
	
	\begin{figure}[]
		\centering
		\includegraphics[width=.6\linewidth]{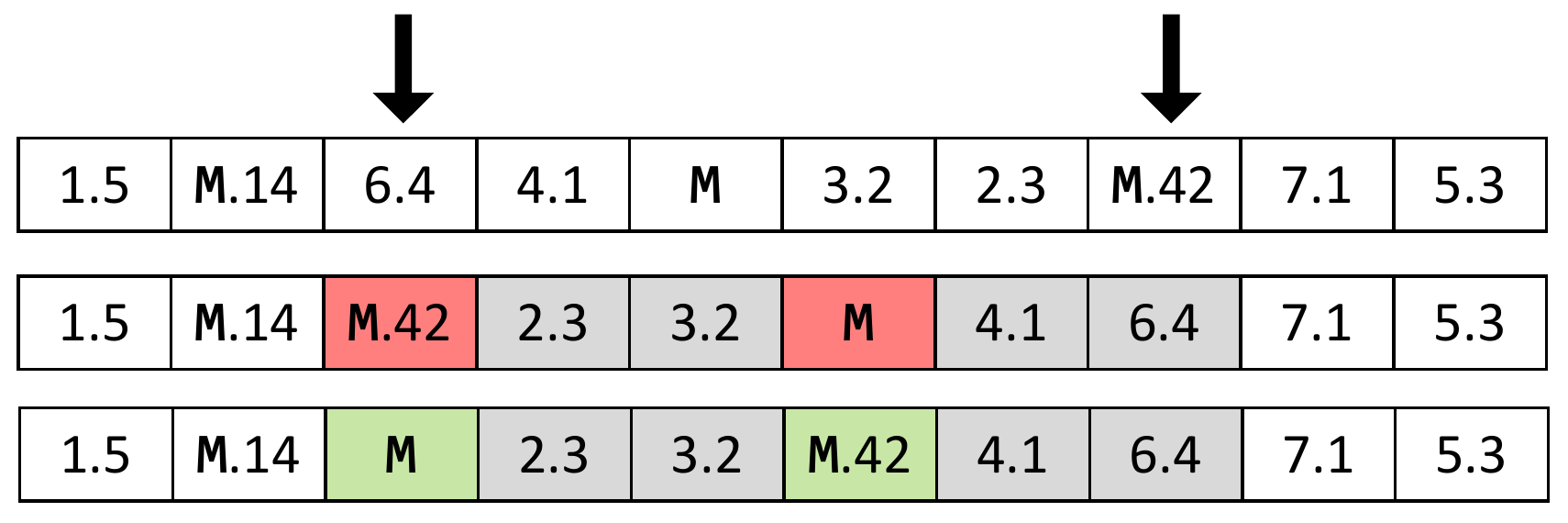}
		\caption{An example of global 2-opt operator.}	
		\label{fig:global_2_opt}
	\end{figure}
	
	The global 2-opt operator selects genes a and b at two different positions in the chromosome, then updates the chromosome by reversing the order of a through b.
	If an odd number of delimiters are included in the selected gene, there is no problem to encode the chromosome after the sequence is inverted.
	If the number of delimiters is even, however, a chromosome is generated in which the divider is successive or the depot cluster is successive.
	In this case, the fractional part of the delimiters is rearranged so that there is no problem encoding the chromosome.
	For example, in Figure \ref{fig:global_2_opt}, the third and eighth positions of a given chromosome are selected (first line) and the chromosome is constructed by rearranging the genes of the corresponding region in the opposite order (second line).
	At this time, there are two delimiters in the selected interval.
	By looking at the second and third genes of the chromosome in the second line, the depot clusters are continuous.
	Therefore, the fractional part existing in the second delimiter of the chromosome is transferred to the third delimiter to construct a chromosome that can be normally encoded.
	If the cost of the chromosome after the global 2-opt is applied is less than the previous cost, the previous information is replaced.
	
	\subsubsection{Local 2-opt}
	
	\begin{figure}[]
		\centering
		\begin{minipage}{\linewidth}
			\centering
			\subfloat{\label{fig:local_2_opt_1}\includegraphics[page=1,width=.6\linewidth]{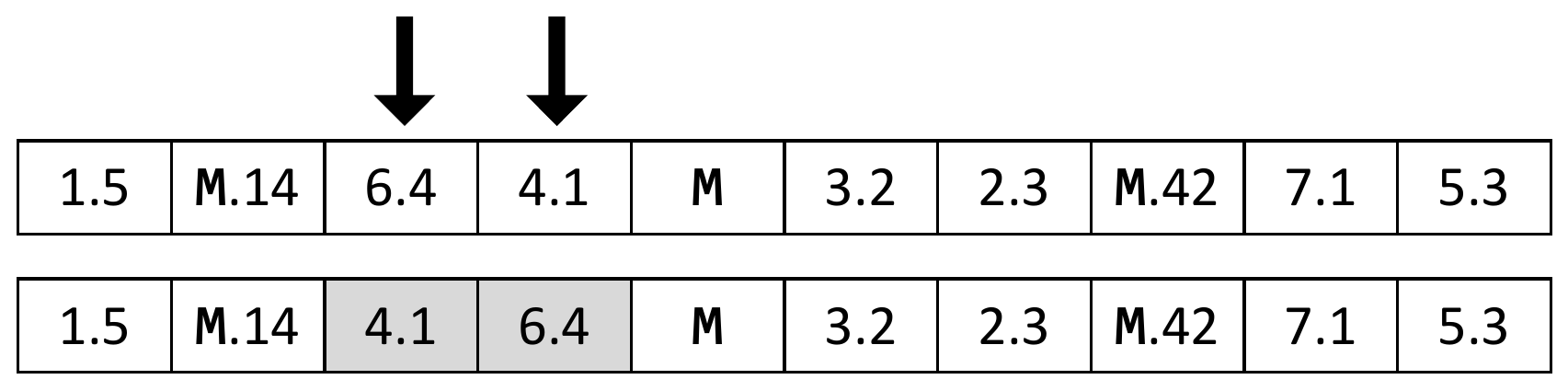}}
		\end{minipage}
		\begin{minipage}{\linewidth}
			\centering
			\subfloat{\label{fig:local_2_opt_2}\includegraphics[page=2,width=.6\linewidth]{local_2_opt}}
		\end{minipage}
		\caption{An example of local 2-opt operator.}	
		\label{fig:local_2_opt}
	\end{figure}
	
	Applying 2-opt by selecting any two genes a and b is similar to the global 2-opt operator, but it does not arbitrarily select the entire chromosome.
	This operator performs 2-opt on the path assigned to each vehicle.
	Taking Figure \ref{fig:local_2_opt} as an example, the given chromosome is divided into paths for vehicles 1 and 2 based on the fifth gene.
	At this point, the upper part of the figure shows a 2-opt example for vehicle 1, and the lower part of the figure shows a 2-opt example for vehicle 2.
	Similar to the global 2-opt operator, the chromosome is replaced when the cost of a chromosome after applying the local 2-opt operator is less than the cost of the previous one.
	
	\subsubsection{Task Swap}
	
	The task swap operator changes position by selecting two different genes.
	Similar to the above operators, the costs of chromosomes are compared and then the better one is used to update the population pool.
	
	\subsubsection{Sample Node Swap}\label{subsubsec:sample_node_swap}
	
	The sample node swap operator updates the chromosome by finding the sample node with the lowest cost for each task cluster.
	The heuristic is applied sequentially from the first gene of a given chromosome.
	If visiting any sample node $ s' $ in the same task cluster is less than visiting the existing sample node $ s $, the chromosome gene information is updated. 	
	One thing to be kept in mind is that if the previous sample node $ s $ has nonempty NIN set $ T^{\textrm{NIN}}_s $, the tasks in $ T^{\textrm{NIN}}_s $ should be checked to determine whether they are covered by any of the sample nodes in the chromosome if they are not included in $ T^{\textrm{NIN}}_{s'} $.	
	%	The Algorithm \ref{alg:sample-node-swap} describes the sample node swap operator in detail.
	%
	%	\begin{algorithm}[t]
	%		\caption{Sample Node Swap}
	%		\begin{algorithmic}[1]
	%			\Procedure{Sample-Node-Swap (chromosome)}{}
	%			\ForEach {gene in chromosome}
	%				\State $ t \leftarrow $ task index from gene
	%				\State $ s \leftarrow $ node from gene
	%				\State $ s_1 \leftarrow $ starting node of edge heading to $ s $
	%				\State $ s_2 \leftarrow $ terminal node of edge from $ s $
	%				\State cost $ \leftarrow c(s_1,s) + c(s,s_2) $
	%				\ForEach {sample node $ s' $ in $ V^k_t \setminus \{s\} $}
	%					\If {any task in $ T^{\textrm{NIN}}_s(s) $ is not reached \\by $ s' $ or other nodes in chromosome}
	%						\State continue
	%					\EndIf
	%					\State cost$ ' \leftarrow c(s_1,s') + c(s',s_2) $
	%					\If {cost$ ' < $ cost}
	%						\State $ s \leftarrow s' $
	%						\State Update gene's sample node information
	%					\EndIf
	%				\EndFor
	%			\EndFor
	%			\RETURN chromosome
	%			\EndProcedure
	%		\end{algorithmic}
	%		\label{alg:sample-node-swap}
	%	\end{algorithm}
	
	\subsection{Population Management and Termination Criteria}
	
	To ensure diversity of the solutions in the GA and not only to enlarge the size of the population, it is necessary to check whether the chromosomes in the population actually have different information.	
	If two chromosomes with the same information in the crossover operator are selected as the parent, the information of the child chromosome is almost unchanged from the information of the parent whatever crossover heuristic is used.
	This means that the convergence rate of the entire GA is slowed down if the chromosome with redundant information in the population is not removed.
	In order to avoid duplication between chromosomes, it is not enough to check whether the gene values of any two chromosomes are sequentially equal, so whether the information actually contained is the same should be checked.
	In this study, we examine the duplication of chromosomes simply by their costs.
	For efficient management, chromosomes are sorted in an ascending order based on the cost for each generation.
	
	The algorithm is terminated when the number of generations reaches the given limit or when the cost of the best chromosome in the pool is not improved for finite consecutive times.

	\subsection{Path Refinement}\label{subsec:path_refine}
	
	As an additional step in the proposed algorithms, this part suggests a process to refine the paths for each vehicle to improve the quality of the solution.
	The output of the sampling based methods for the Dubins TSP with neighborhoods is obtained through a limited number of crudely discretized samples in a 3-dimensional space.
	In other words, even if the optimal solution is obtained for a given roadmap, there is some quality difference from the actual optimum solution with the given conditions.
	Therefore, to reduce the difference, a local optimization is applied to the outputs from the proposed algorithms.
	When a local optimization is performed, the parameters are optimized in the continuous state space while it is assumed that the newly refined path follows the sequence of visiting each neighborhood from the given solution.
	The improvement of the solution quality through this step might be limited since the visiting sequence of the vehicle does not change.
	However, the global optimal solution of the instance can be obtained if the quality of the given solution from the algorithms above is sufficiently high.
	In addition, there is an advantage in that the total calculation time can be greatly reduced compared to simply increasing the total number of samples in the instance.
	
	Path refinement proceeds as follows. Since the output of the previous step may not have sample nodes for all tasks due to the NIN, the order and state of entry into the neighborhood of each task are sequentially generated based on the path of each vehicle in the given solution.
	Considering the states of the depots and the terminals, the total number of states to be optimized is $ n+2m' $ where n is the number of tasks, and $ 2m' $ correspond to the number of initial and final states for each vehicle assigned at least a single task.
	For a local optimization of each task state, the constraints of the state to be optimized and the neighboring states are required.
	Each is optimized in the direction of decreasing cost where the neighboring states are fixed.\footnote{The `fmincon' function in MATLAB is used for the local optimization.}
	Similarly, the depot state is optimized with the next state fixed and the terminal state with the previous state fixed.
	Optimization is repeated in an alternating order as follows: odd-numbered states are optimized while others are fixed and then even-numbered states are optimized, and the iteration repeats until the cost of vehicle converges.
	In the simulation, iteration was performed until the difference between the previous cost and the next cost was less than 0.01\%.

	\section{Numerical Experiments}\label{sec:num_exp}
	
	In this section, we discuss the computational results of the formulated problem to address the following questions:
	\begin{enumerate}
		\item How does the performance differ depending on whether NIN and path refinement are applied?
		\item How large a problem instance can be handled practically?
		\item When each vehicle is heterogeneous, is it possible to generate mission-effective tours considering the characteristics of each vehicle?
		\item How do the tours change with the change of the coefficient $ \alpha $ in the objective function?
	\end{enumerate}
	
	There are a number of parameters that can vary the characteristics of problem instance, therefore, first a numerical simulation is performed by changing the number of vehicles and sample nodes in each task cluster while fixing the values of other parameters.
	After that, the results of the simulation are reported in each section while the parameters that are considered to be important are changed.
	Detailed parameters are listed in Table \ref{tab:param} with their values.
	
	\begin{table}[]
		\centering
		\caption{Parameters for simulation. Bold values are for default values.}
		\label{tab:param}
		\begin{tabular}{c|c}
			\hthickline
			\textbf{Parameters} & \textbf{Value} \\ \hline\hline
			Number of vehicles & \{\textbf{1},2,3,\textbf{4}\}\\ \hline
			Number of sample nodes in each task & \{1,2,3,4,\textbf{5},10,20,30,40,50\} \\ \hline
			\begin{tabular}{c} Task sensing range of vehicle \\ (radius of a task neighborhood for vehicle) (m) \end{tabular}
			& \{100, \textbf{150}, 200\} \\ \hline
			\hspace*{-0.2cm}\begin{tabular}{c} Vehicle's velocity and \\ minimum turning radius $ (\sfrac{m}{s}, m) $ \end{tabular} & \hspace*{-0.3cm} \begin{tabular}{c}\{(50, 65.9), \\(60, 94.8), \\\textbf{(70, 129.1)}\} \end{tabular} \\ \hline
			Depot location (m, m) & \hspace*{-0.3cm} \begin{tabular}{l}\{(110, 230), (1800, 2100), \\ (200, 1500), (1700, 1000)\} \end{tabular} \\ \hline
			Vehicle load factor & 4 \\ \hline
			Metric of $ c_{s,s'} $ (Cost from $ s $ to $ s' $) & \{\textbf{length}, time\} \\\hline
			$ \alpha $ (Objective function coefficient) & \{0, \textbf{0.5}, 1\} \\ \hthickline
		\end{tabular}
	\end{table}
	
	Since one of the main purposes of this paper was to analyze the characteristics of the GHMDATSP, we varied the size of the problem by changing the number of vehicles and sample nodes while the number of tasks was fixed for ease of analysis.
	The instances to be solved are generated from one of the TSPLIB instances called \textit{bays29}, which contains 29 targets in a 2-dimensional region.
	Adding $ m $ depots and terminals given m vehicles, the total number of clusters is 29+2$ m $ for each instance.
	Sample nodes are randomly generated in the circular neighborhoods of each task with a radius of $ 150m $, and the heading directions in each sample node are also given as random.
	The sample nodes in the depot and the terminal are located at fixed positions, and also the headings are randomly given.
	Every vehicle has its unique sample nodes for each cluster, and no nodes are shared between different vehicles.
	%	The default speed of the vehicle is assumed to be $ 75 m/s $ and the load factor is set as 4, so the minimum turning radius of the vehicle is $ 148.2m $.
	%	Except the instances in Section \ref{subsec:hetero}, the dynamic characteristics of each vehicle are the same as the default values.	
	Unless otherwise mentioned, instances are solved by applying the NIN, and the value of the objective function coefficient $ \alpha $ is 0.5.
	
	We compare the following methods:
	\begin{enumerate}[label*=\arabic*.]
		\item EA-noNIN, EA-NIN, EA-NIN-PR are the methods based on the MILP implementation.
		\begin{enumerate}[label*=\arabic*.]
			\item EA-noNIN: without considering the NIN.
			\item EA-NIN: applying NIN.
			\item EA-NIN-PR: applying path refinement on the solution of EA-NIN.
		\end{enumerate}
		\item MA-noNIN, MA-NIN, MA-NIN-PR are the methods based on the memetic algorithm.
		\begin{enumerate}[label*=\arabic*.]
			\item MA-noNIN: without considering the NIN.
			\item MA-NIN: applying NIN.
			\item MA-NIN-PR: applying path refinement on the solution of MA-NIN.
		\end{enumerate}
		\item OOD is the heuristics by Obermeyer et al. \cite{obermeyer2012sampling}. OOD is a sampling based method, which transforms the problem into the form of the ATSP and solve it using the LKH heuristic. It only can handle a single vehicle problem.
		\item ZCXP is the heuristics by Zhang et al. \cite{zhang2014memetic}. ZCXP is based on the memetic algorithm and solves the problem in the continuous domain in terms of the location and heading for each task.
	\end{enumerate}	
	
	The capital letters `EA' for the methods using MILP implementation are borrowed from the word `exact algorithm', and the detailed description of the corresponding methods is provided in the appendix.
	All of the MILP based methods were performed on a PC with Intel(R) Xeon(R) CPU E5-2687W v4 @ 3.00GHz and 64.0GB RAM using Gurobi 7.5.1 as a MILP solver.
%	The EA based approaches were implemented in Python environment using the lazy callback functionality of Gurobi.
	The calculation time for each instance is reported in seconds, and the algorithm has an upper limit of 7,200 seconds.
	The results from the other mehtods were performed on a PC with Intel(R) CPU i7-6700K and 16.0GB RAM.
	The MA based mehtods were implemented in the C\#, and the other heuristics (OOD and ZCXP) were implemented in MATLAB environment.\footnote{The C code of the LKH heuristic was mex-compiled to use in MATLAB environment.}
	Before comparing the performance of the above mentioned methods quantitatively, the difference of the results from applying the NIN and the path refinement process is explained qualitatively as follows.
	
	\subsection{NIN and Path Refinement}\label{subsec:NINandPR}
	
	\begin{figure*}[]
		\centering
		\captionsetup{justification=centering}
		\subfloat[Tour result from EA-noNIN.]{\label{fig:res1_noNIN}\includegraphics[width=.32\linewidth]{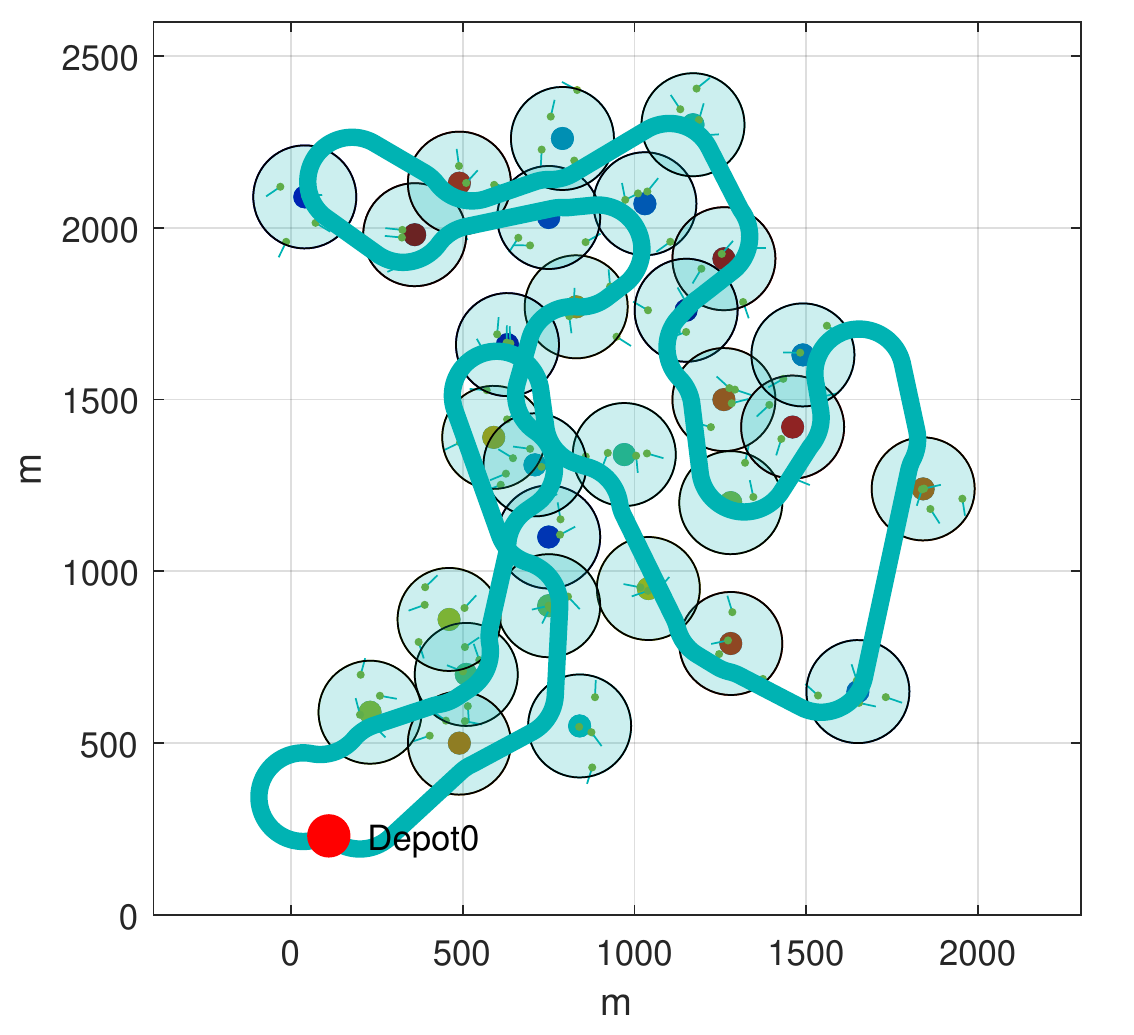}}
		\subfloat[Tour result from EA-NIN.]{\label{fig:res1_NIN}\includegraphics[width=.32\linewidth]{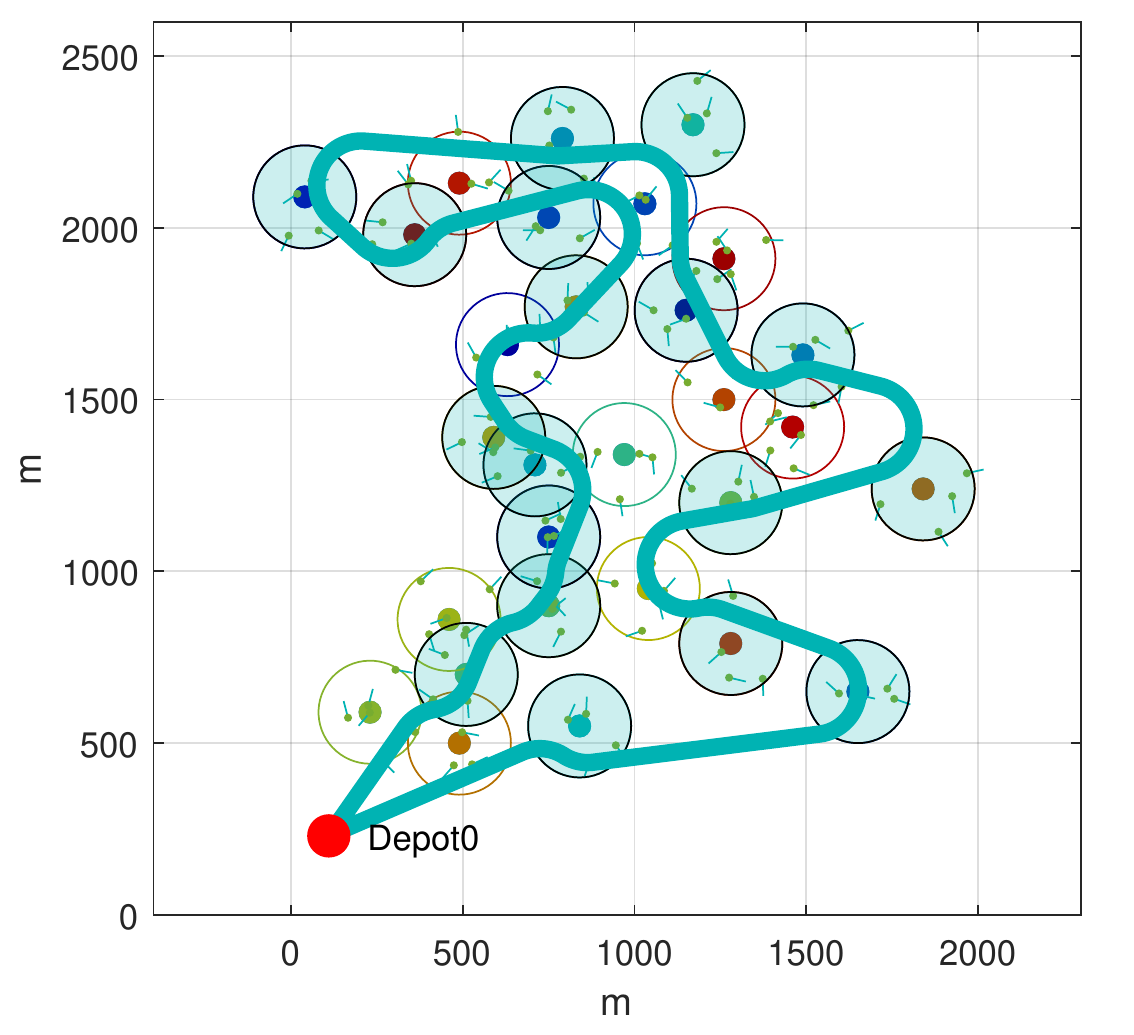}}
		\subfloat[Tour result from EA-NIN-PR.]{\label{fig:res1_NINPR}\includegraphics[width=.32\linewidth]{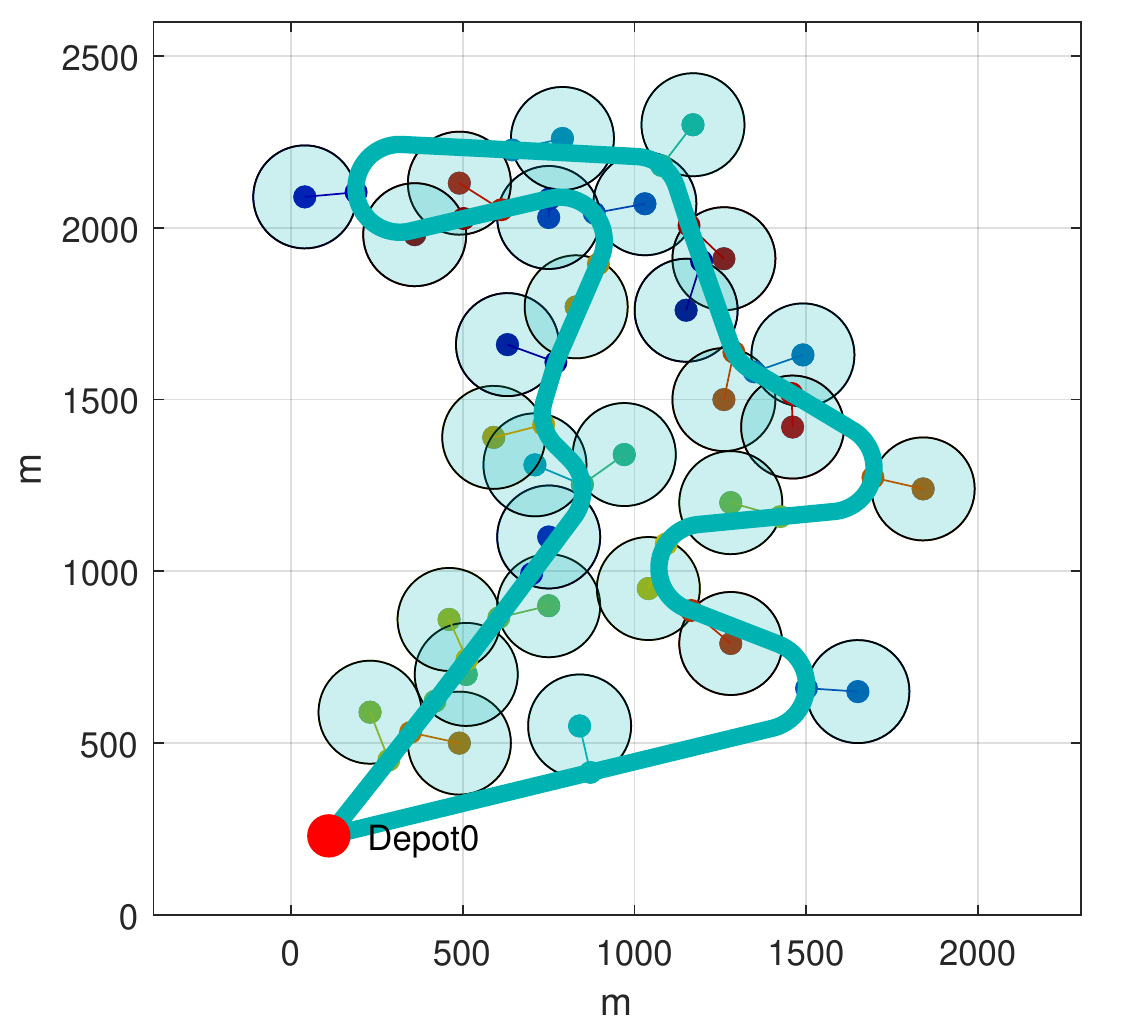}}
		\caption{Tours given a single vehicle, varying the method. Velocity: 70m/s.}		
		\label{fig:tourRes1}
	\end{figure*}
	
	\begin{figure*}[]
		\centering
		\captionsetup{justification=centering}
		\subfloat[Tour result from MA-noNIN.]{\label{fig:res1_vs11}\includegraphics[width=.32\linewidth]{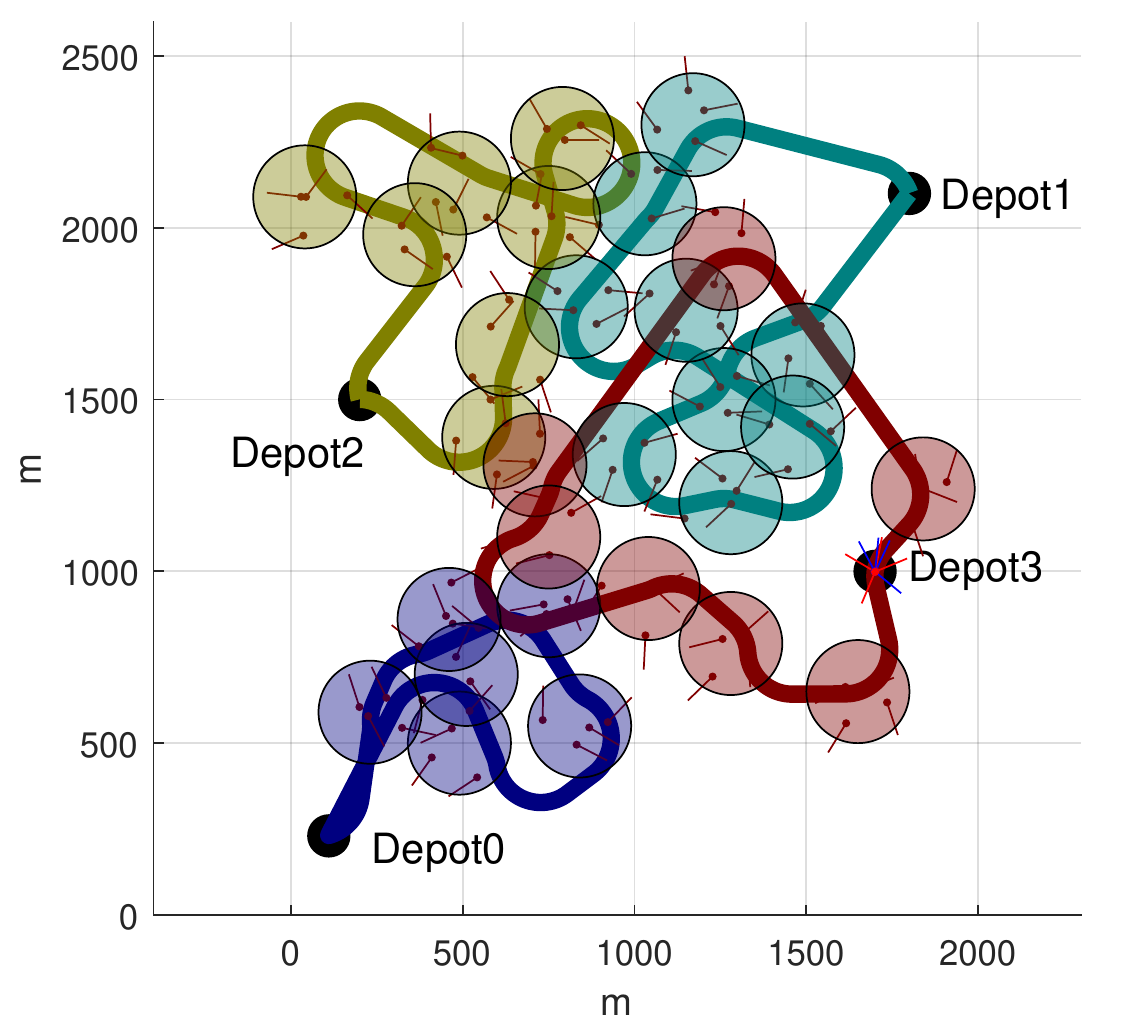}}
		\subfloat[Tour result from MA-NIN.]{\label{fig:res1_vs15}\includegraphics[width=.32\linewidth]{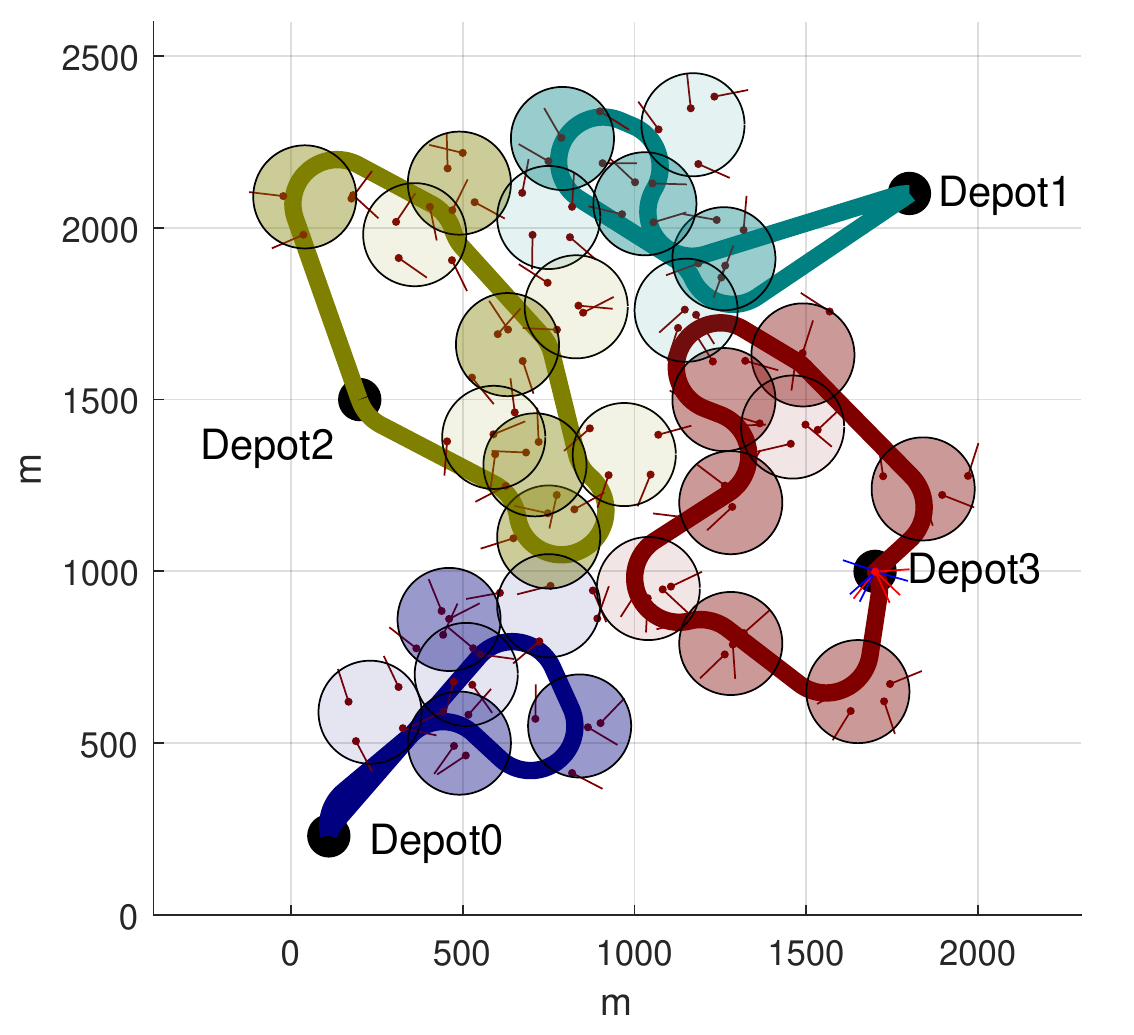}}
		\subfloat[Tour result from MA-NIN-PR.]{\label{fig:res1_vs41}\includegraphics[width=.32\linewidth]{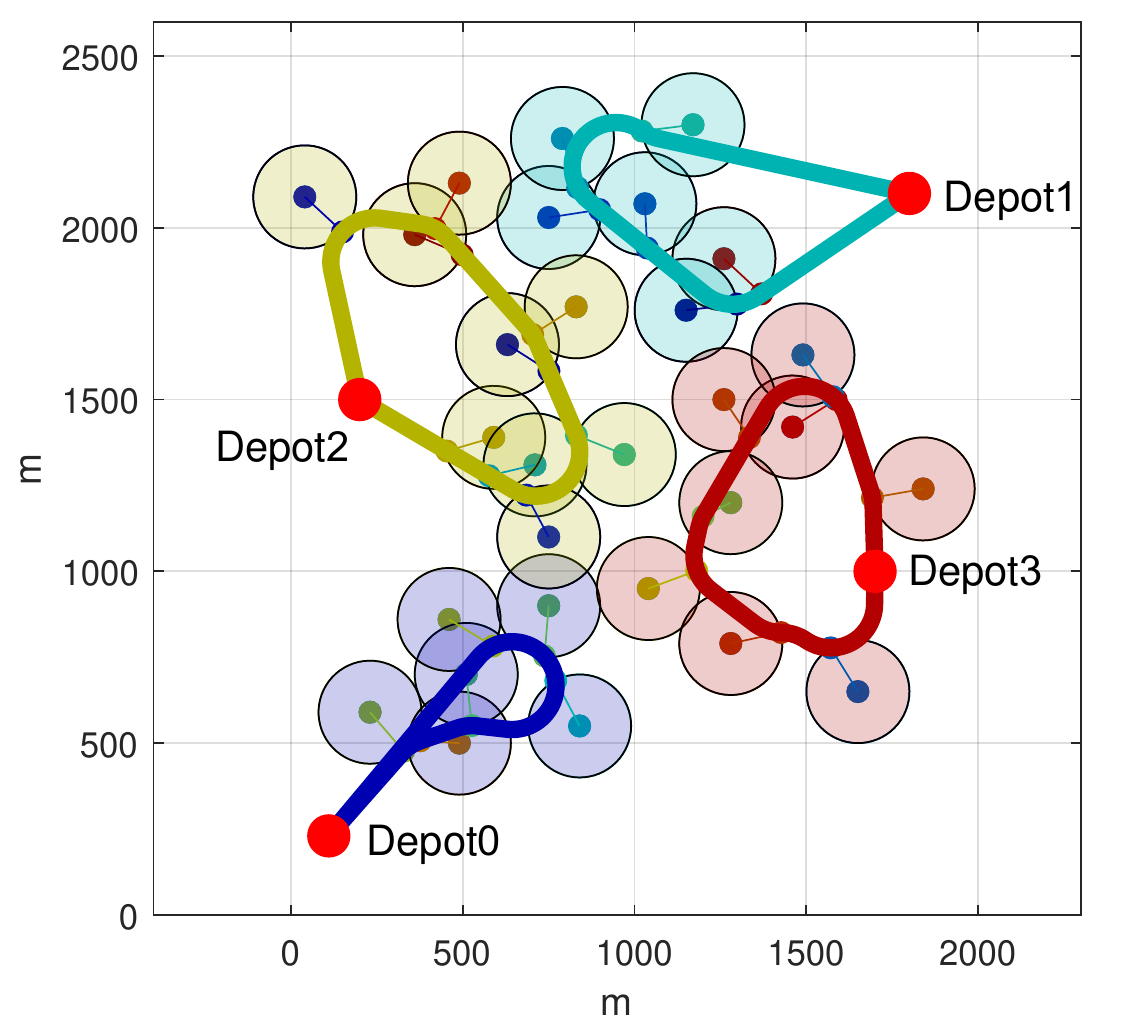}}
		\caption{Tours given four vehicles, varying the method. Velocity: 70m/s.}		
		\label{fig:tourRes2}
	\end{figure*}
	
	The closer the tasks are located, and the larger the turning radius of the vehicle, the more different the characteristics of the solution of the DTSPN from the general ETSP.
	The dramatic change in the results of each method is shown in this section given the big turning radius (velocity: 70m/s, radius: 129.1m).
	Figure \ref{fig:tourRes1} shows the results for an instance where a single vehicle tours and five sample nodes are created in each cluster.
	Each result of Figure \ref{fig:res1_noNIN} and Figure \ref{fig:res1_NIN} is the optimal solution for the given instance through the solver and the results generated from EA-noNIN, EA-NIN, and EA-NIN-PR are shown in order from left to right.
	
	Although it is optimal in Figure \ref{fig:res1_noNIN}, to visit one of the nodes of every task exactly, the result made a detour more than necessary since the number samples is small.
	The result of Figure \ref{fig:res1_NIN} is quite encouraging in that the detour of the path is significantly reduced via indirect visits compared to the Figure \ref{fig:res1_noNIN}.
	Each indirectly visited task has only a border and is not colored.
	The result in Figure \ref {fig:res1_NINPR} is obtained by applying the path refinement process to the result of Figure \ref{fig:res1_NIN}, resulting in a cost reduction of about 15\%.
	It is not the global optimal solution for a given instance, but it can be seen that a satisfactory result can be obtained through an effective heuristic even with a small number of samples.
	Figure \ref{fig:tourRes2} shows the results for an instance where four vehicles are available, and the results from MA-noNIN, MA-NIN, and MA-NIN-PR are shown in order from left to right.
	Similar to the results in Figure \ref{fig:tourRes1}, the cost is reduced when NIN and path refinement is applied.

	\subsection{Comparative experiments}\label{subsec:CompExp}
	
	\begin{figure*}[]
		\centering
		\captionsetup{justification=centering}
		\subfloat[Comparison of suggested methods.]{\label{fig:objComp1}\includegraphics[width=.45\linewidth]{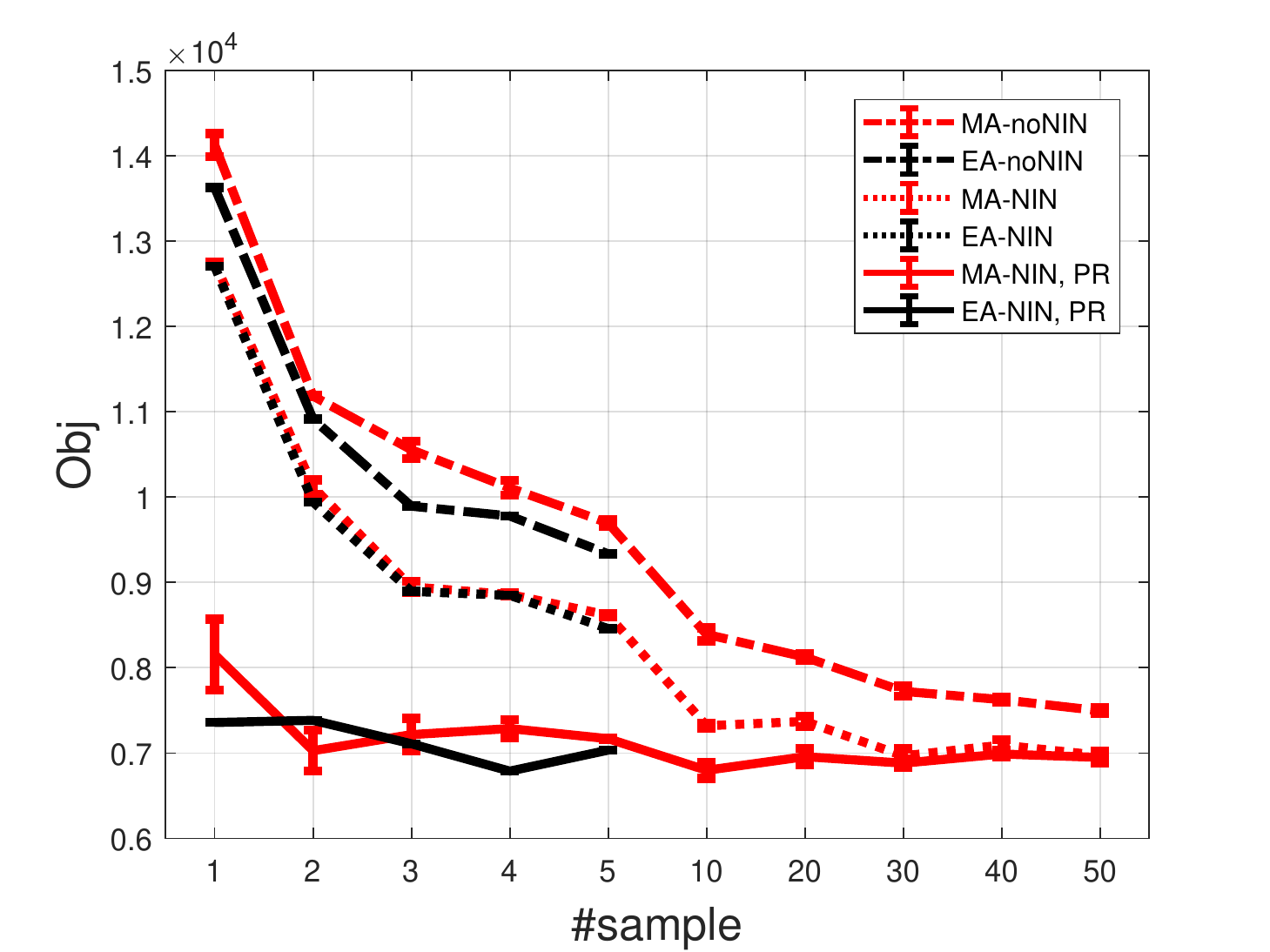}}
		\subfloat[Comparison with other methods.]{\label{fig:objComp2}\includegraphics[width=.45\linewidth]{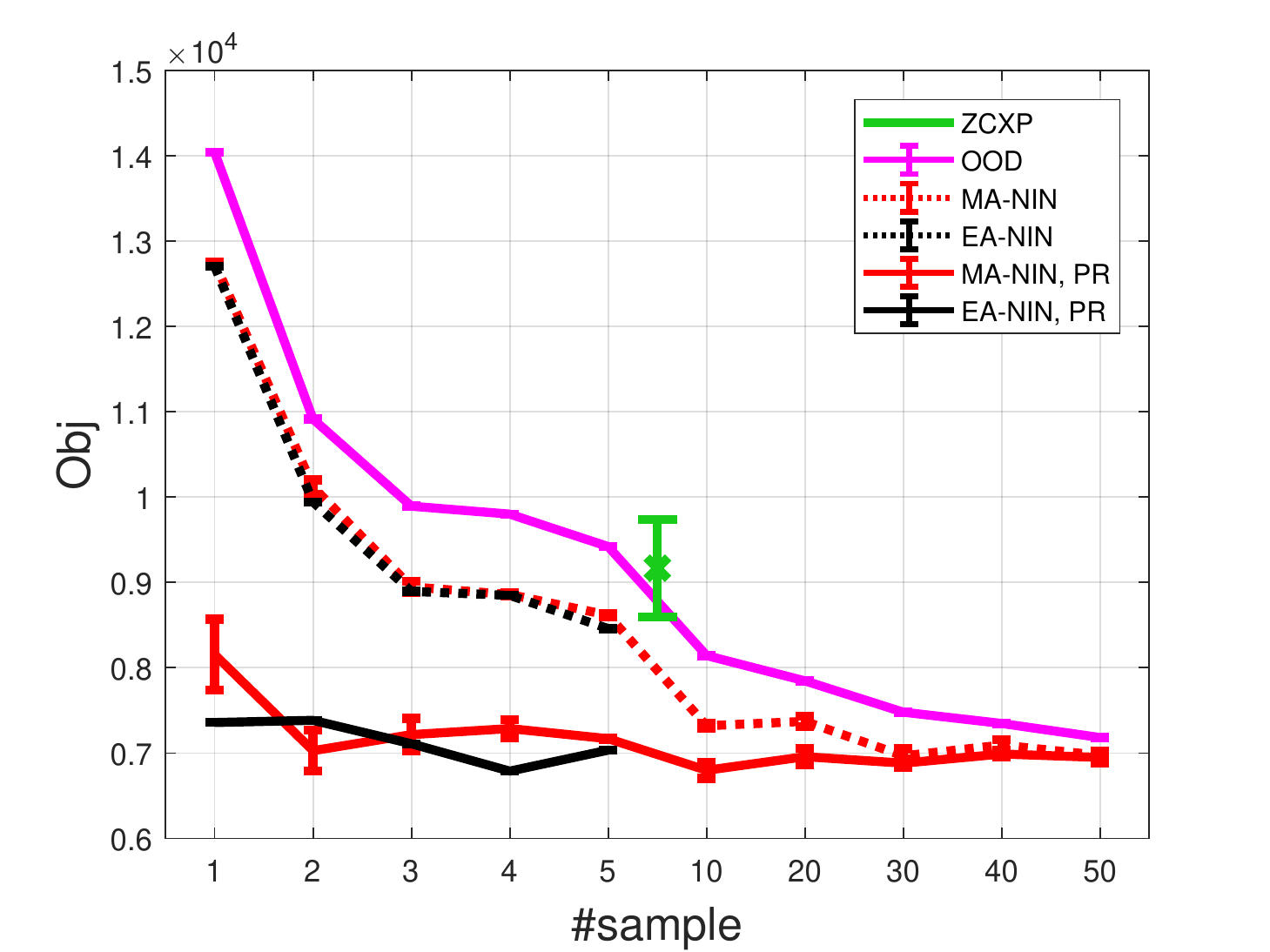}}
		\caption{Comparison of objective values for each method. Single vehicle, velocity: 50m/s.}		
		\label{fig:objComp}
	\end{figure*}
	
	\begin{figure}[]
		\centering
		\includegraphics[width=.5\linewidth]{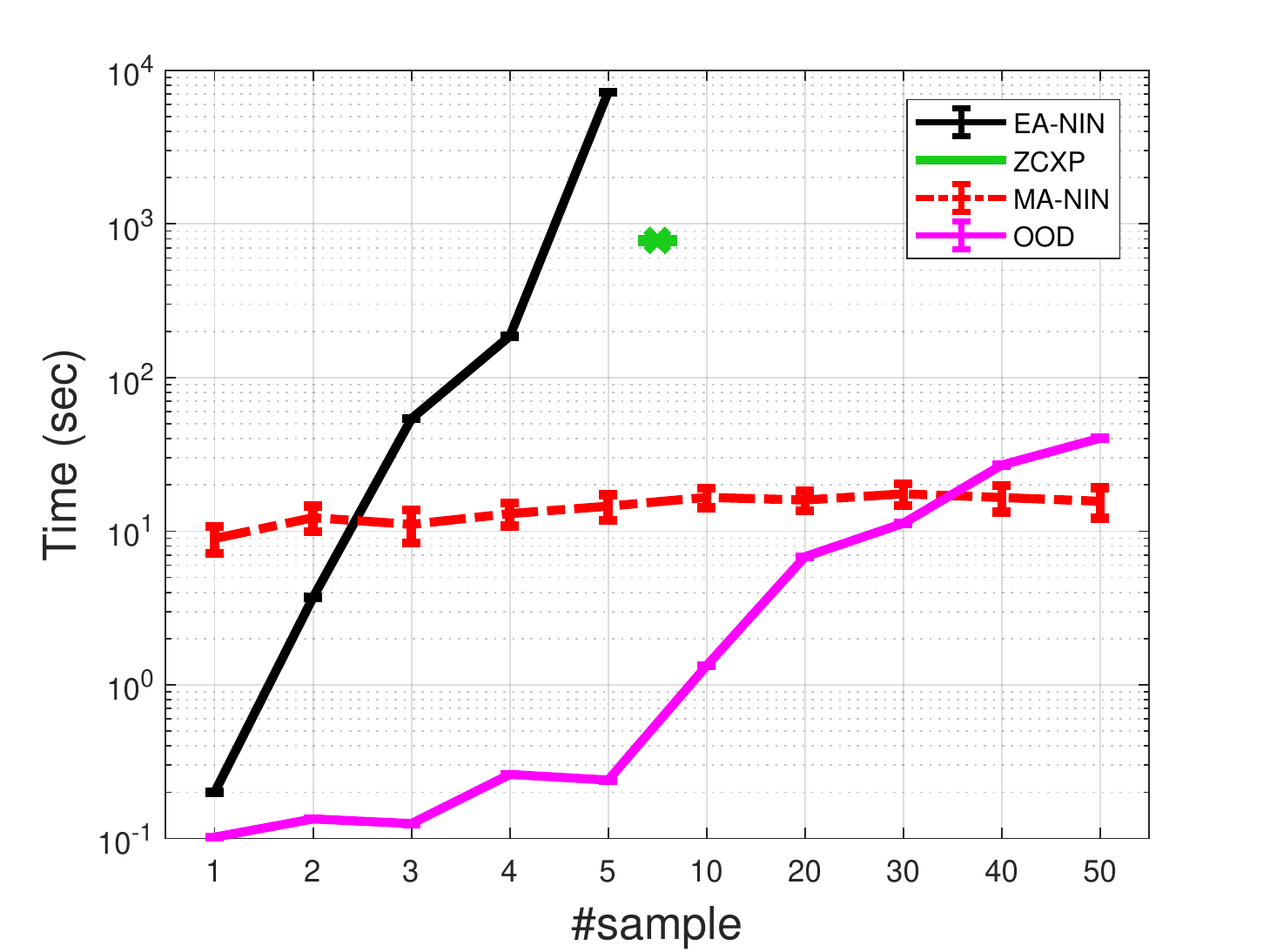}
		\caption{Comparison of computational times for each method.}
		\label{fig:timeComp}
	\end{figure}
	
	Figure \ref{fig:objComp} shows the comparison of the proposed methods and other methods quantitatively with the mean and the standard deviation of the objective values through the error bar.
	The instance used in Figure \ref{fig:objComp} is generated assuming a single vehicle with a speed set to 50m/s.
	The results of the memetic algorithm based methods (MA-noNIN, MA-NIN, MA-NIN-PR, and ZCXP) are generated ten times for the same instance.
	The results of the MILP based methods and the OOD are generated once since they generate almost the same outputs for the same instance.
	In addition, the change of the output according to the number of sample nodes cannot be confirmed from the ZCXP because it is not a sampling based method.
	In order to ensure convergence of the objective value, the maximum number of generations for the ZCXP is fixed to 10,000 times.
	Although not shown in the figure, the MA methods proposed in the paper satisfied the terminate condition after about several tens to hundreds of generations, and for the ZCXP it was also confirmed that the objective values almost converged after hundreds of generations.
	
	Figure \ref{fig:objComp1} shows the objective values of the proposed methods with the number of sample nodes in the cluster being changed.
	For the MILP implementation, the maximum number of sample nodes was limited to five due to its complexity.
	The performance of the solution is improved in the order of MA-noNIN, EA-noNIN, MA-NIN, EA-NIN, MA-NIN-PR and EA-NIN-PR.
	Though it depends on the number of sample nodes, it is shown that the objective value can be reduced by up to 50\% by the path refinement.
	Furthermore, it can be seen that each method gradually converges to the near-optimal as the number of sample nodes increases.
	Figure \ref{fig:objComp2} shows a comparison of the results from the proposed methods and other methods.
	The performance of the proposed methods is superior to the ZCXP and OOD.
	The reason why the ZCXP shows poor performance is that the change of visiting sequence depends only on the crossover and mutation operators, which makes it difficult for the chromosome to escape from the local optimals.
	The OOD shows slightly better performance than MA-noNIN but poorer than the other proposed methods.
	As in Figure \ref{fig:objComp1}, the result of OOD shows that the objective value gradually decreases as the number of sample nodes increases.
	
	Figure \ref{fig:timeComp} shows the comparison of the computational time of each method.
	Since the difference of each method is large, the time axis is represented by log scale.
	Because the time complexity of the path refinement is $ O(n) $ and the time required for cost convergence was around one to two seconds, the computational time results for the path refinement process were excluded from the comparison.
	In the case of the MILP implementation, the computational time is the most sensitive to an increase in the number of sample nodes.
	The computational time of the memetic algorithm does not change much since the terminal condition is constant.
	Also there is no other factor except the increase in the calculation time of the sample node swap operator (Section \ref{subsubsec:sample_node_swap}) as the number of samples increases.
	A direct comparison with other methods is difficult since the ZCXP was run in the MATLAB environment, but it differs by about 50 times from the proposed MA based methods which are implemented in C\#.
	The computational time of the OOD increases steeply with the increase of the number of samples since it uses the LKH heuristic whose time complexity is known to be $ O(n^{2.2}) $ \cite{helsgaun2015solving}.
	
	Although not shown in the figure, we performed simulations with an increasing number of vehicles from one to four, changing the vehicle speed from 50m/s to 60 and 70m/s, and found that the tendency of the objective value and computational time follows the results in Figures \ref{fig:objComp1} and \ref{fig:objComp2}.
	Table \ref{tab:result1} shows the results of the above methods with varying the number of vehicles and sample nodes.
	The MILP solver could not obtain the optimal solution for most of the cases due to the complexity of the problem, and the suboptimal results are shaded in red in Table \ref{tab:result1}.
	As mentioned above, the number of samples is limited up to five for the MILP implementation and the OOD is the heuristic for a single vehicle.
	The instances not calculated are shaded in gray in Table \ref{tab:result1}.

	\subsection{Heterogeneous Motion Constraints}\label{subsec:hetero}
	
	\begin{figure}[]
		\centering
		\includegraphics[width=.7\linewidth,trim={0.8cm 0.5cm 0.8cm 1.2cm}]{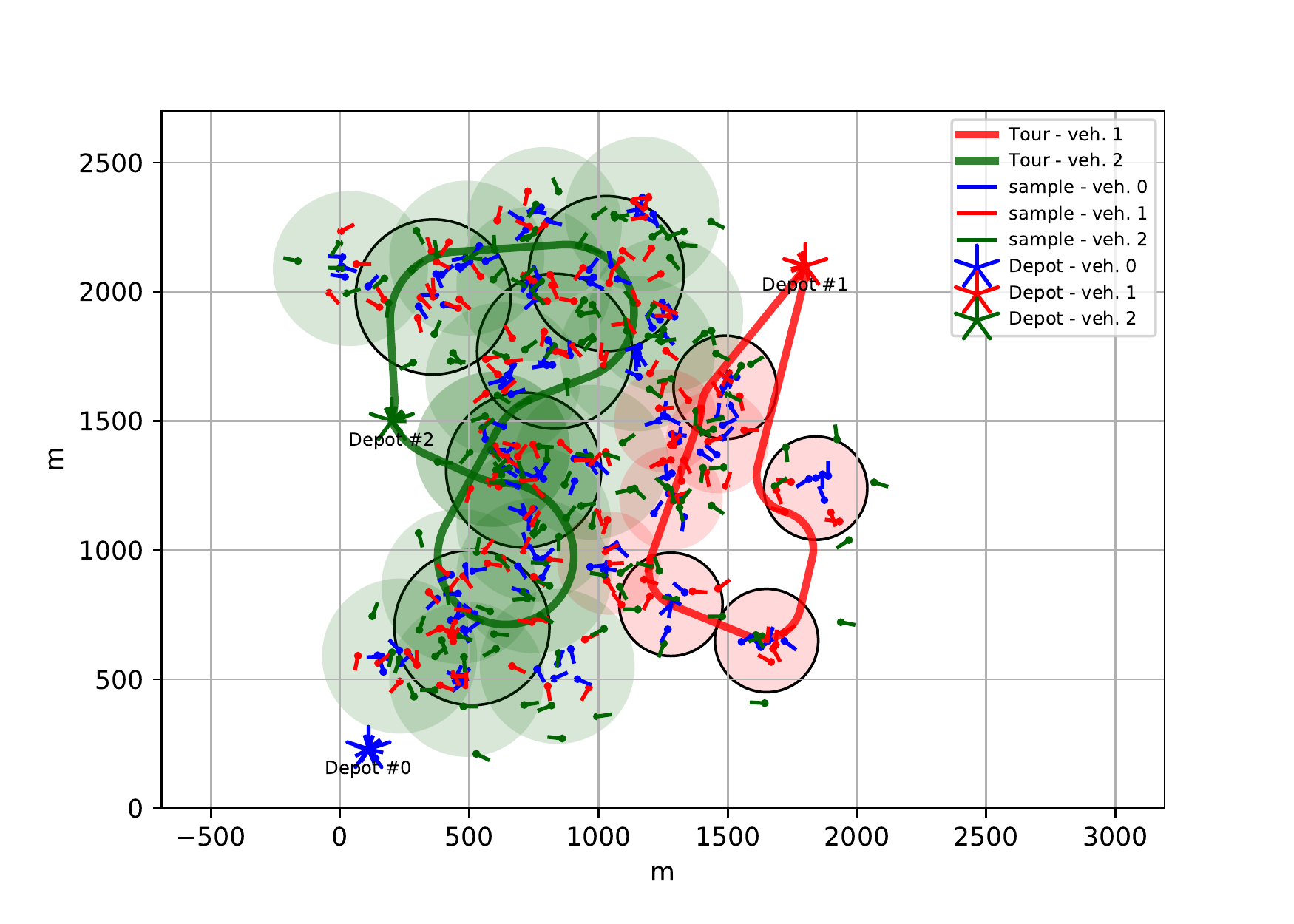}
		\caption{Optimal tours for a heterogeneous fleet.}	
		\label{fig:res2}
	\end{figure}
	
	To show how the algorithm handles a fleet of structurally heterogeneous vehicles, an instance with different vehicle characteristics are given to be solved.
	Three vehicles are given as vehicles \#0, \#1, and \#2, with the velocity given as 50m/s, 75m/s, and 100m/s and the sensing range given as 100m, 150m, and 200m respectively, and the cost metric is chosen as `time'.
	The dynamics of each vehicle is different and follows the values in Table \ref{tab:param}.
	Vehicle \#0 is slow and the remote sensing coverage is small; \#2 is fast and the coverage is large; and the characteristics of \#1 are about halfway between them.
	Figure \ref{fig:res2} shows the optimal solution of the instance using the EA-NIN method, and the number of nodes per cluster are given as 5.
	Vehicle \#2 handles most of the tasks in the region because of its large coverage and agility, and then \#1 handles the remaining tasks.
	In the solution, no path is assigned to \#0 due to its poor performance.

	\subsection{Various Objective Functions}\label{subsec:obj}
	
	\begin{figure*}[]
		\centering
		\captionsetup{justification=centering}
		\subfloat[Instance - (\#v, \#s) : (4, 5), $ \alpha=1 $]{\label{fig:res3_1}\includegraphics[width=.32\linewidth]{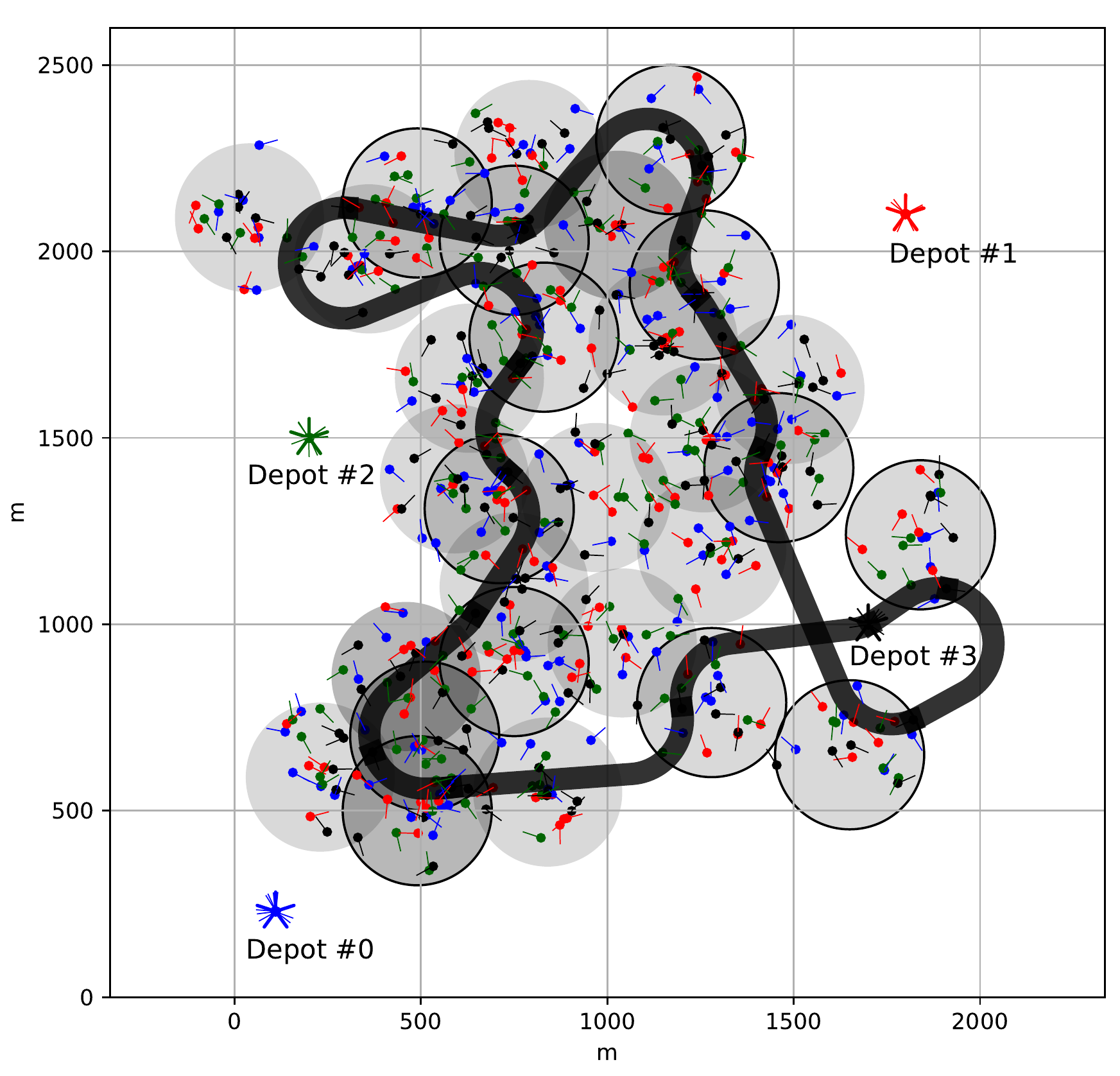}}
		\subfloat[Instance - (\#v, \#s) : (4, 5), $ \alpha=0.5 $]{\label{fig:res3_2}\includegraphics[width=.32\linewidth]{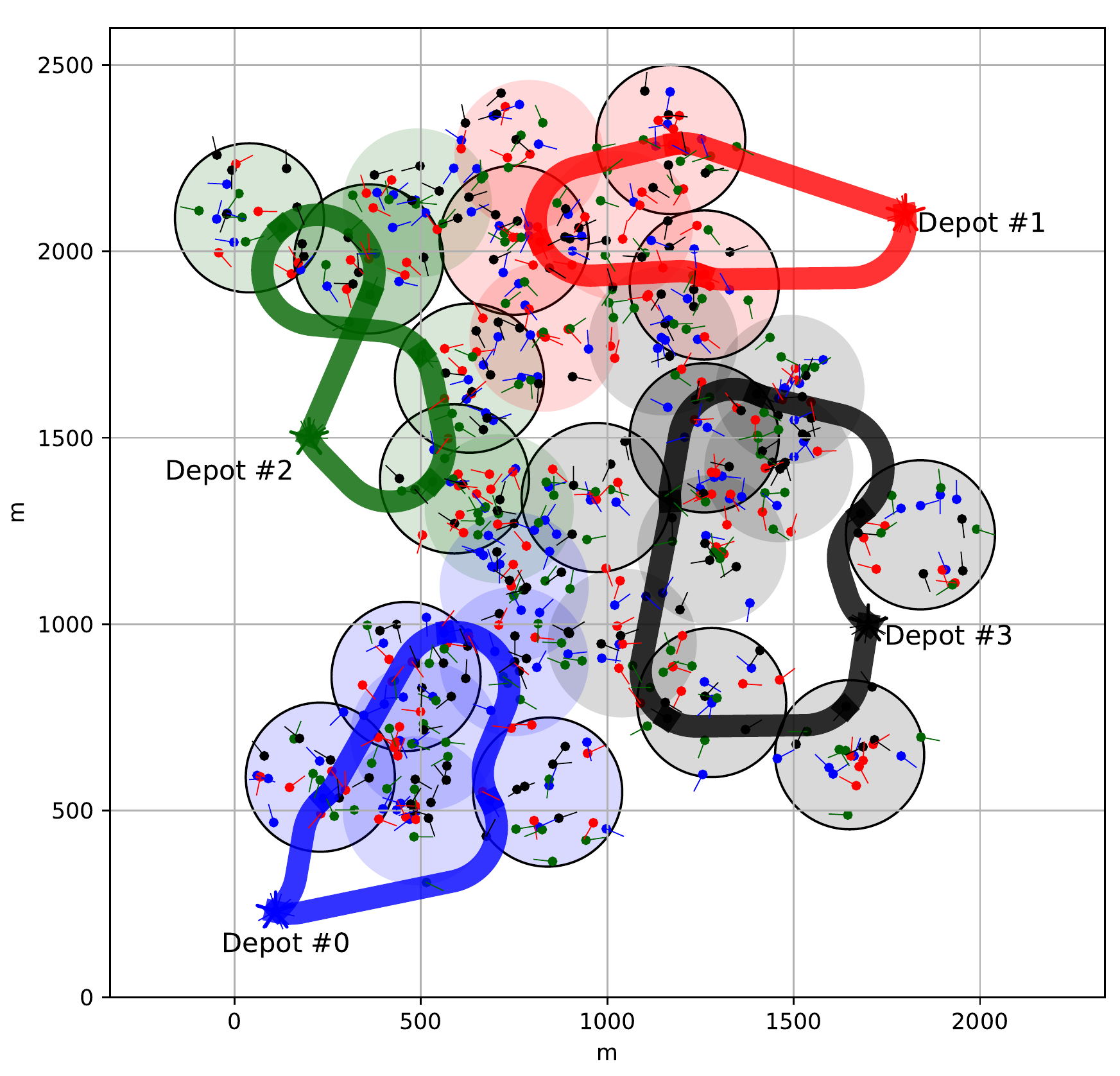}}
		\subfloat[Instance - (\#v, \#s) : (4, 5), $ \alpha=0 $]{\label{fig:res3_3}\includegraphics[width=.32\linewidth]{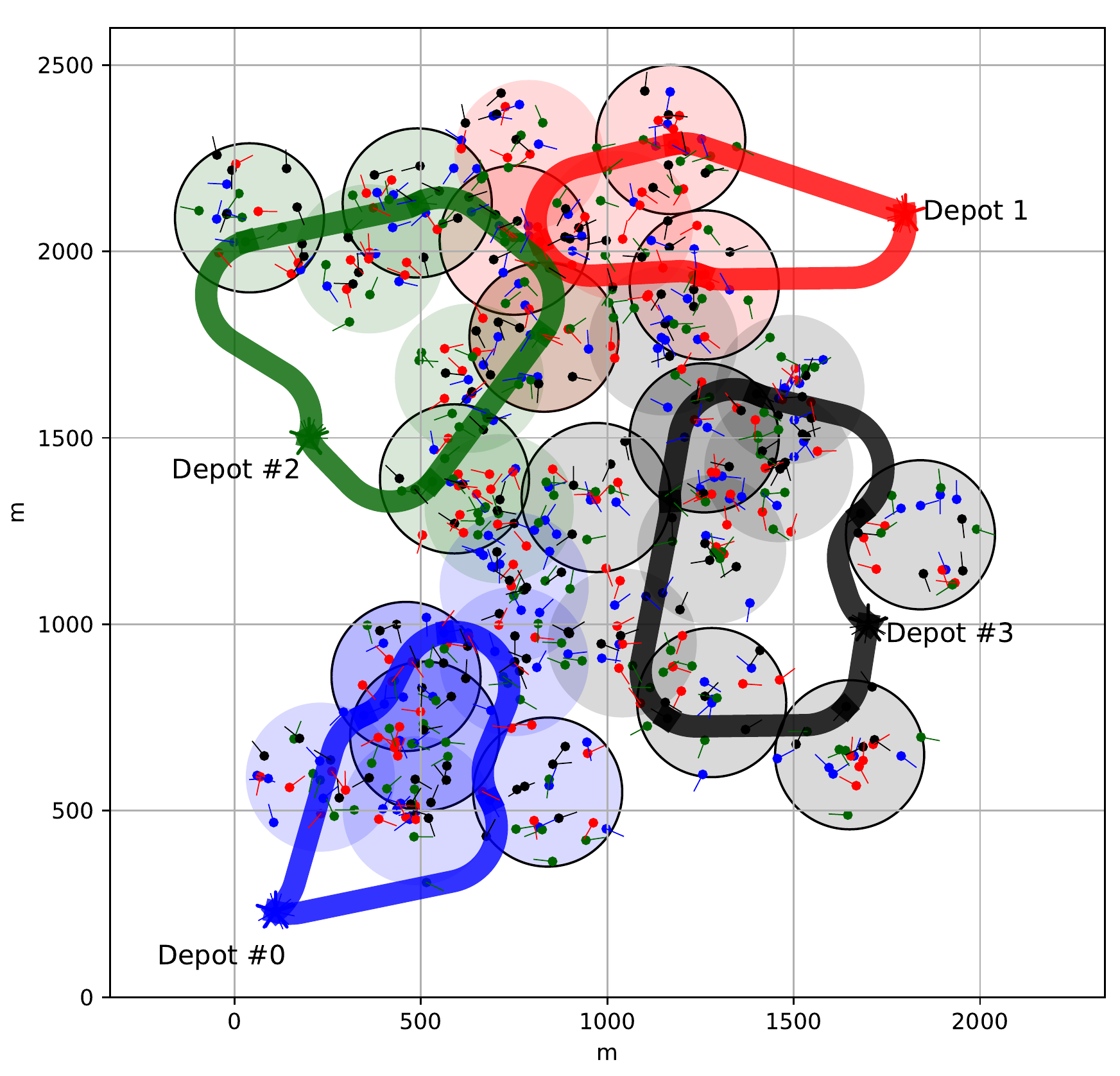}}
		\caption{Optimal tours with varying $ \alpha $.}		
		\label{fig:res3}
	\end{figure*}
	
	Using the default parameters and (\#v, \#s) as (4,5), the coefficient value $ \alpha $ in eq. \eqref{eq:obj} is varied to check how the feature of results changes.
	Figures \ref{fig:res3_1} and \ref{fig:res3_2} show the optimal results of given instances when the value of $ \alpha $ is 1 and 0, respectively.
	As mentioned, the formulated problem with $ \alpha=1 $ is a typical \textit{min} problem, and if $ \alpha=0 $ then the problem becomes a \textit{min-max} problem.
	The sum of tour costs in Figure \ref{fig:res3_1} is 7,744, which is smaller than the instance in Figure \ref{fig:res3_2} (9,511), but all of the tasks are assigned to the vehicle \#3.
	The phenomenon in which the task assignment is concentrated only on a few vehicles often occurs in the multiple TSP where the objective function is given to minimize all the mission costs.
	In case of $ \alpha = 0 $, the interesting result is that the objective value (which is the same as the maximum tour cost among the vehicles) is 2,706, which is the same as the instance where $ \alpha $ was 0.5.
	But the total sum of costs given $ \alpha = 0 $ is 9,943, which is larger than the case of cost 9,512 when $ \alpha $ is 0.5.
		
	\begin{table}[htbp]
		\centering
		\caption{Computational results, velocity set to 50m/s.}
		\begin{adjustbox}{max width=\textwidth}
		\begin{tabular}{c|c|crcrcrccrcrc}
			\cmidrule{4-14}    \multicolumn{2}{c}{} &       & \multicolumn{6}{c}{Objective cost}            &       & \multicolumn{4}{c}{Time (sec)} \\
			\cmidrule{4-9}\cmidrule{11-14}    \multicolumn{1}{p{1.75em}}{\#v} & \multicolumn{1}{p{1.75em}}{\#s} &       & \multicolumn{1}{p{8em}}{EA-NIN, gap(\%)} & \multicolumn{1}{p{3.94em}}{MA-NIN} & \multicolumn{1}{p{3.94em}}{EA-NIN, PR} & \multicolumn{1}{p{4.2em}}{MA-NIN, PR} & \multicolumn{1}{p{3.625em}}{OOD} & \multicolumn{1}{p{3.125em}}{ZCXP} &       & \multicolumn{1}{p{3em}}{EA} & \multicolumn{1}{p{3em}}{MA} & \multicolumn{1}{p{3em}}{OOD} & \multicolumn{1}{p{2.75em}}{ZCXP} \\
			\midrule
			\multirow{10}[2]{*}{1} & 1     &       & \multicolumn{1}{c}{12702.4} & 12751.4 & \multicolumn{1}{c}{7358.6} & 7623.7 & \multicolumn{1}{c}{14037.3} & \multirow{10}[2]{*}{9166.0} &       & \multicolumn{1}{c}{0.2} & 9.0   & \multicolumn{1}{c}{0.1} & \multirow{10}[2]{*}{780.8} \\
			& 2     &       & \multicolumn{1}{c}{9942.0} & 10118.1 & \multicolumn{1}{c}{7383.9} & 7351.3 & \multicolumn{1}{c}{10913.1} &       &       & \multicolumn{1}{c}{3.7} & 12.2  & \multicolumn{1}{c}{0.1} &  \\
			& 3     &       & \multicolumn{1}{c}{8894.4} & 8946.8 & \multicolumn{1}{c}{7109.5} & 7157.3 & \multicolumn{1}{c}{9892.7} &       &       & \multicolumn{1}{c}{185.9} & 11.1  & \multicolumn{1}{c}{0.1} &  \\
			& 4     &       & \multicolumn{1}{c}{8848.2} & 8858.6 & \multicolumn{1}{c}{6788.9} & 6651.1 & \multicolumn{1}{c}{9800.0} &       &       & \multicolumn{1}{c}{53.9} & 13.0  & \multicolumn{1}{c}{0.3} &  \\
			& 5     &       & \multicolumn{1}{c}{\cellcolor[rgb]{ 1,  .757,  .757}8458.8 (2.96\%)} & 8616.1 & \multicolumn{1}{c}{7035.0} & 7085.3 & \multicolumn{1}{c}{9419.9} &       &       & \multicolumn{1}{c}{\cellcolor[rgb]{ 1,  .757,  .757}7200} & 14.5  & \multicolumn{1}{c}{0.2} &  \\
			& 10    &       & \cellcolor[rgb]{ .949,  .949,  .949} & 7320.3 & \cellcolor[rgb]{ .949,  .949,  .949} & 6650.1 & \multicolumn{1}{c}{8143.0} &       &       & \cellcolor[rgb]{ .949,  .949,  .949} & 16.6  & \multicolumn{1}{c}{1.3} &  \\
			& 20    &       & \cellcolor[rgb]{ .949,  .949,  .949} & 7371.9 & \cellcolor[rgb]{ .949,  .949,  .949} & 6688.9 & \multicolumn{1}{c}{7846.3} &       &       & \cellcolor[rgb]{ .949,  .949,  .949} & 15.9  & \multicolumn{1}{c}{6.8} &  \\
			& 30    &       & \cellcolor[rgb]{ .949,  .949,  .949} & 6967.6 & \cellcolor[rgb]{ .949,  .949,  .949} & 6671.0 & \multicolumn{1}{c}{7479.5} &       &       & \cellcolor[rgb]{ .949,  .949,  .949} & 17.5  & \multicolumn{1}{c}{11.3} &  \\
			& 40    &       & \cellcolor[rgb]{ .949,  .949,  .949} & 7093.6 & \cellcolor[rgb]{ .949,  .949,  .949} & 6639.2 & \multicolumn{1}{c}{7347.4} &       &       & \cellcolor[rgb]{ .949,  .949,  .949} & 16.6  & \multicolumn{1}{c}{26.9} &  \\
			& 50    &       & \cellcolor[rgb]{ .949,  .949,  .949} & 6973.9 & \cellcolor[rgb]{ .949,  .949,  .949} & 6667.8 & \multicolumn{1}{c}{7182.4} &       &       & \cellcolor[rgb]{ .949,  .949,  .949} & 15.7  & \multicolumn{1}{c}{40.3} &  \\
			\cmidrule{1-2}\cmidrule{4-9}\cmidrule{11-14}    \multirow{10}[2]{*}{2} & 1     &       & \multicolumn{1}{c}{6255.1} & 6383.1 & \multicolumn{1}{c}{4885.9} & 4652.4 & \cellcolor[rgb]{ .949,  .949,  .949} & \multirow{10}[2]{*}{5414.8} &       & \multicolumn{1}{c}{41.4} & 12.2  & \cellcolor[rgb]{ .949,  .949,  .949} & \multirow{10}[2]{*}{819.3} \\
			& 2     &       & \multicolumn{1}{c}{5590.8} & 5673.0 & \multicolumn{1}{c}{4379.8} & 4262.9 & \cellcolor[rgb]{ .949,  .949,  .949} &       &       & \multicolumn{1}{c}{1080.9} & 12.8  & \cellcolor[rgb]{ .949,  .949,  .949} &  \\
			& 3     &       & \multicolumn{1}{c}{\cellcolor[rgb]{ 1,  .757,  .757}5498.6 (9.19\%)} & 5515.0 & \multicolumn{1}{c}{4566.1} & 4418.8 & \cellcolor[rgb]{ .949,  .949,  .949} &       &       & \multicolumn{1}{c}{\cellcolor[rgb]{ 1,  .757,  .757}7200} & 12.6  & \cellcolor[rgb]{ .949,  .949,  .949} &  \\
			& 4     &       & \multicolumn{1}{c}{4729.5} & 4812.3 & \multicolumn{1}{c}{4216.2} & 4254.1 & \cellcolor[rgb]{ .949,  .949,  .949} &       &       & \multicolumn{1}{c}{2485.0} & 14.1  & \cellcolor[rgb]{ .949,  .949,  .949} &  \\
			& 5     &       & \multicolumn{1}{c}{\cellcolor[rgb]{ 1,  .757,  .757}5447.3 (18.49\%)} & 5021.9 & \multicolumn{1}{c}{4682.8} & 4223.3 & \cellcolor[rgb]{ .949,  .949,  .949} &       &       & \multicolumn{1}{c}{\cellcolor[rgb]{ 1,  .757,  .757}7200} & 16.4  & \cellcolor[rgb]{ .949,  .949,  .949} &  \\
			& 10    &       & \cellcolor[rgb]{ .949,  .949,  .949} & 4561.3 & \cellcolor[rgb]{ .949,  .949,  .949} & 4144.6 & \cellcolor[rgb]{ .949,  .949,  .949} &       &       & \cellcolor[rgb]{ .949,  .949,  .949} & 15.9  & \cellcolor[rgb]{ .949,  .949,  .949} &  \\
			& 20    &       & \cellcolor[rgb]{ .949,  .949,  .949} & 4384.7 & \cellcolor[rgb]{ .949,  .949,  .949} & 4095.1 & \cellcolor[rgb]{ .949,  .949,  .949} &       &       & \cellcolor[rgb]{ .949,  .949,  .949} & 18.7  & \cellcolor[rgb]{ .949,  .949,  .949} &  \\
			& 30    &       & \cellcolor[rgb]{ .949,  .949,  .949} & 4225.9 & \cellcolor[rgb]{ .949,  .949,  .949} & 4012.0 & \cellcolor[rgb]{ .949,  .949,  .949} &       &       & \cellcolor[rgb]{ .949,  .949,  .949} & 17.5  & \cellcolor[rgb]{ .949,  .949,  .949} &  \\
			& 40    &       & \cellcolor[rgb]{ .949,  .949,  .949} & 4200.5 & \cellcolor[rgb]{ .949,  .949,  .949} & 4041.2 & \cellcolor[rgb]{ .949,  .949,  .949} &       &       & \cellcolor[rgb]{ .949,  .949,  .949} & 17.5  & \cellcolor[rgb]{ .949,  .949,  .949} &  \\
			& 50    &       & \cellcolor[rgb]{ .949,  .949,  .949} & 4196.6 & \cellcolor[rgb]{ .949,  .949,  .949} & 4069.6 & \cellcolor[rgb]{ .949,  .949,  .949} &       &       & \cellcolor[rgb]{ .949,  .949,  .949} & 20.1  & \cellcolor[rgb]{ .949,  .949,  .949} &  \\
			\cmidrule{1-2}\cmidrule{4-9}\cmidrule{11-14}    \multirow{10}[2]{*}{3} & 1     &       & \multicolumn{1}{c}{4420.6} & 4582.8 & \multicolumn{1}{c}{2937.4} & 3093.0 & \cellcolor[rgb]{ .949,  .949,  .949} & \multirow{10}[2]{*}{3873.3} &       & \multicolumn{1}{c}{44.3} & 13.0  & \cellcolor[rgb]{ .949,  .949,  .949} & \multirow{10}[2]{*}{809.7} \\
			& 2     &       & \multicolumn{1}{c}{\cellcolor[rgb]{ 1,  .757,  .757}4158.4 (7.14\%)} & 4199.0 & \multicolumn{1}{c}{3236.2} & 3255.7 & \cellcolor[rgb]{ .949,  .949,  .949} &       &       & \multicolumn{1}{c}{\cellcolor[rgb]{ 1,  .757,  .757}7200} & 15.6  & \cellcolor[rgb]{ .949,  .949,  .949} &  \\
			& 3     &       & \multicolumn{1}{c}{\cellcolor[rgb]{ 1,  .757,  .757}3766.3 (8.40\%)} & 3718.2 & \multicolumn{1}{c}{3271.5} & 3054.7 & \cellcolor[rgb]{ .949,  .949,  .949} &       &       & \multicolumn{1}{c}{\cellcolor[rgb]{ 1,  .757,  .757}7200} & 15.1  & \cellcolor[rgb]{ .949,  .949,  .949} &  \\
			& 4     &       & \multicolumn{1}{c}{\cellcolor[rgb]{ 1,  .757,  .757}3693.9 (12.26\%)} & 3668.1 & \multicolumn{1}{c}{3165.4} & 3114.3 & \cellcolor[rgb]{ .949,  .949,  .949} &       &       & \multicolumn{1}{c}{\cellcolor[rgb]{ 1,  .757,  .757}7200} & 15.5  & \cellcolor[rgb]{ .949,  .949,  .949} &  \\
			& 5     &       & \multicolumn{1}{c}{\cellcolor[rgb]{ 1,  .757,  .757}3534.7 (17.27\%)} & 3537.5 & \multicolumn{1}{c}{3030.0} & 3054.3 & \cellcolor[rgb]{ .949,  .949,  .949} &       &       & \multicolumn{1}{c}{\cellcolor[rgb]{ 1,  .757,  .757}7200} & 16.9  & \cellcolor[rgb]{ .949,  .949,  .949} &  \\
			& 10    &       & \cellcolor[rgb]{ .949,  .949,  .949} & 3311.5 & \cellcolor[rgb]{ .949,  .949,  .949} & 2982.3 & \cellcolor[rgb]{ .949,  .949,  .949} &       &       & \cellcolor[rgb]{ .949,  .949,  .949} & 17.5  & \cellcolor[rgb]{ .949,  .949,  .949} &  \\
			& 20    &       & \cellcolor[rgb]{ .949,  .949,  .949} & 3153.7 & \cellcolor[rgb]{ .949,  .949,  .949} & 2953.2 & \cellcolor[rgb]{ .949,  .949,  .949} &       &       & \cellcolor[rgb]{ .949,  .949,  .949} & 17.5  & \cellcolor[rgb]{ .949,  .949,  .949} &  \\
			& 30    &       & \cellcolor[rgb]{ .949,  .949,  .949} & 3187.4 & \cellcolor[rgb]{ .949,  .949,  .949} & 3005.4 & \cellcolor[rgb]{ .949,  .949,  .949} &       &       & \cellcolor[rgb]{ .949,  .949,  .949} & 15.7  & \cellcolor[rgb]{ .949,  .949,  .949} &  \\
			& 40    &       & \cellcolor[rgb]{ .949,  .949,  .949} & 3104.4 & \cellcolor[rgb]{ .949,  .949,  .949} & 2956.1 & \cellcolor[rgb]{ .949,  .949,  .949} &       &       & \cellcolor[rgb]{ .949,  .949,  .949} & 21.4  & \cellcolor[rgb]{ .949,  .949,  .949} &  \\
			& 50    &       & \cellcolor[rgb]{ .949,  .949,  .949} & 3151.5 & \cellcolor[rgb]{ .949,  .949,  .949} & 3001.3 & \cellcolor[rgb]{ .949,  .949,  .949} &       &       & \cellcolor[rgb]{ .949,  .949,  .949} & 19.9  & \cellcolor[rgb]{ .949,  .949,  .949} &  \\
			\cmidrule{1-2}\cmidrule{4-9}\cmidrule{11-14}    \multirow{10}[2]{*}{4} & 1     &       & \multicolumn{1}{c}{3477.0} & 3541.2 & \multicolumn{1}{c}{2549.1} & 2595.4 & \cellcolor[rgb]{ .949,  .949,  .949} & \multirow{10}[2]{*}{2982.4} &       & \multicolumn{1}{c}{1198.9} & 13.3  & \cellcolor[rgb]{ .949,  .949,  .949} & \multirow{10}[2]{*}{890.4} \\
			& 2     &       & \multicolumn{1}{c}{\cellcolor[rgb]{ 1,  .757,  .757}3059.1 (5.68\%)} & 3076.7 & \multicolumn{1}{c}{2238.2} & 2237.4 & \cellcolor[rgb]{ .949,  .949,  .949} &       &       & \multicolumn{1}{c}{\cellcolor[rgb]{ 1,  .757,  .757}7200} & 17.0  & \cellcolor[rgb]{ .949,  .949,  .949} &  \\
			& 3     &       & \multicolumn{1}{c}{\cellcolor[rgb]{ 1,  .757,  .757}2696.8 (8.04\%)} & 2767.7 & \multicolumn{1}{c}{2118.6} & 2210.9 & \cellcolor[rgb]{ .949,  .949,  .949} &       &       & \multicolumn{1}{c}{\cellcolor[rgb]{ 1,  .757,  .757}7200} & 17.6  & \cellcolor[rgb]{ .949,  .949,  .949} &  \\
			& 4     &       & \multicolumn{1}{c}{\cellcolor[rgb]{ 1,  .757,  .757}2632.2 (8.21\%)} & 2658.4 & \multicolumn{1}{c}{2172.1} & 2119.0 & \cellcolor[rgb]{ .949,  .949,  .949} &       &       & \multicolumn{1}{c}{\cellcolor[rgb]{ 1,  .757,  .757}7200} & 16.6  & \cellcolor[rgb]{ .949,  .949,  .949} &  \\
			& 5     &       & \multicolumn{1}{c}{\cellcolor[rgb]{ 1,  .757,  .757}2622.7 (15.86\%)} & 2535.0 & \multicolumn{1}{c}{2171.0} & 2092.0 & \cellcolor[rgb]{ .949,  .949,  .949} &       &       & \multicolumn{1}{c}{\cellcolor[rgb]{ 1,  .757,  .757}7200} & 18.9  & \cellcolor[rgb]{ .949,  .949,  .949} &  \\
			& 10    &       & \cellcolor[rgb]{ .949,  .949,  .949} & 2385.9 & \cellcolor[rgb]{ .949,  .949,  .949} & 2148.1 & \cellcolor[rgb]{ .949,  .949,  .949} &       &       & \cellcolor[rgb]{ .949,  .949,  .949} & 19.6  & \cellcolor[rgb]{ .949,  .949,  .949} &  \\
			& 20    &       & \cellcolor[rgb]{ .949,  .949,  .949} & 2272.8 & \cellcolor[rgb]{ .949,  .949,  .949} & 2093.5 & \cellcolor[rgb]{ .949,  .949,  .949} &       &       & \cellcolor[rgb]{ .949,  .949,  .949} & 18.4  & \cellcolor[rgb]{ .949,  .949,  .949} &  \\
			& 30    &       & \cellcolor[rgb]{ .949,  .949,  .949} & 2201.0 & \cellcolor[rgb]{ .949,  .949,  .949} & 2042.2 & \cellcolor[rgb]{ .949,  .949,  .949} &       &       & \cellcolor[rgb]{ .949,  .949,  .949} & 20.5  & \cellcolor[rgb]{ .949,  .949,  .949} &  \\
			& 40    &       & \cellcolor[rgb]{ .949,  .949,  .949} & 2202.7 & \cellcolor[rgb]{ .949,  .949,  .949} & 2043.0 & \cellcolor[rgb]{ .949,  .949,  .949} &       &       & \cellcolor[rgb]{ .949,  .949,  .949} & 20.9  & \cellcolor[rgb]{ .949,  .949,  .949} &  \\
			& 50    &       & \cellcolor[rgb]{ .949,  .949,  .949} & 2205.8 & \cellcolor[rgb]{ .949,  .949,  .949} & 2090.3 & \cellcolor[rgb]{ .949,  .949,  .949} &       &       & \cellcolor[rgb]{ .949,  .949,  .949} & 20.5  & \cellcolor[rgb]{ .949,  .949,  .949} &  \\
			\bottomrule
		\end{tabular}%
		\end{adjustbox}
		\label{tab:result1}%
	\end{table}%

	\section{Conclusion}\label{sec:conc}
	
	In this paper, the methods and procedure of memetic algorithm based path generation was presented as variants of the TSP, which is called the generalized heterogeneous multiple depot asymmetric traveling salesmen problem (GHMDATSP), which arises in the context of achieving a mission of remote surveillance using a fleet of Dubins vehicles.
	A mixed-integer linear programming formulation was proposed to solve instances of the GHMDATSP.
	The suggested procedure incorporates a robust tour improvement heuristics to fit the GHMDATSP into the classic genetic algorithm.	
%	For the MILP implementation based methods, a customized branch-and-cut algorithm was developed using the proposed formulation.	
	To enhance the performance of the solutions, first we used the concept called the necessarily intersecting neighborhood (NIN), which generated effective paths, especially when the tasks were densely located, and second we used the local optimization based path refinement process to find the best visiting location and heading of every task region.
	In order to observe the characteristics of the GHMDATSP with Dubins vehicles and to verify the efficiency of the proposed methods, a wide class of simulation was performed on the instances generated from a standard library with diverse variations of the key parameters.
	The solutions were obtained for up to 4 vehicles and 29 tasks with different numbers of sample nodes for each cluster.
	Numerical simulations verified that the proposed methods were superior to the other two state-of-the-art methods in terms of the performance and the computational time.
	And the comparison with the results from MILP implementation based methods shows the proposed methods can
	
	The proposed methods can be further extended to incorporate the limited maximum travel distance for each vehicle.
	Another possible direction for the research is to reduce the number of tasks by adapting the geometric sensor cover problem.

	\section*{Appendix - MILP Implementation}\label{sec:append}
	
	In this section, we describe the major parts of the MILP implementation used to solve the GHMDATSP.
	The optimal solution can be obtained by providing the formulation developed in Section \ref{sec:background} to an off-the-shelf commercial MILP solver.
	The problem with the subtour elimination constraints, however, is that the number of these constraints increases exponentially as the number of task increases.
	It is known that the number of subtour elimination constraints exceeds $ 10^{15} $ for a normal TSP with 50 tasks; therefore generating all subtour elimination constraints is impossible.
	
	To handle this issue, the constraints in eq. \eqref{eq:subelim} are generated step by step whenever it is needed, which is called the \textit{row generation}.
	Generally, row generation is applied to the constraints that do not occur frequently during the execution of a branch-and-cut algorithm.
	At the beginning of the solving process, the constraints are relaxed from the original formulation.
	Whenever the solver gets a feasible solution of the relaxed problem, the callback procedure is called which checks whether the solution violates any of the constraints in Eq. \eqref{eq:subelim}.
	If the tasks assigned to each vehicle are connected and satisfy the subtour elimination constraints, the solution is regarded as a true solution for the given problem and it is accepted.
	Otherwise, the solution is abandoned and every subtour which does not include the depot and terminal clusters is put as the ingredient of new constraints.
	After these constraints are added to the formulation, the branch-and-cut algorithm continues to solve the problem.
	This process is known to solve the traditional traveling salesman problem and its variants in an effective manner.	
	The separation algorithm, which finds every subtour in the feasible solution of the relaxed problem, is provided in Algorithm \ref{alg:subtour}.
	
	\begin{algorithm}[t]
		\caption{Separation algorithm - Find subtours}
		\begin{algorithmic}[1]
			\Procedure{FindSubtour ($ \textbf{x}^*, \textbf{y}^*, \textbf{y}^{\textrm{NIN*}} $)}{}
			\State $ unvisited = \{1,\cdots,n\}$
			\For{$ k := 1 $ to $ m $}
			\State $ s_1 \leftarrow $ Depot sample node of vehicle $ k $ (param: $ \textbf{y} $)
			\State $ s_2 \leftarrow $ Direct successor of $ s_1 $ (param: $ \textbf{x} $)
			\While{}
			\If {$ s_2 $ belongs to a terminal cluster}
			\State break \textbf{while} loop
			\EndIf
			\State $ t \leftarrow $ task index which contains $ s_2 $
			\State Remove $ t $ in $ unvisited $
			\If {$ T^{\textrm{NIN}}_{s_2} \neq \emptyset $ (param: $ \textbf{y}^{\textrm{NIN*}} $)}
			\State Remove NIN tasks of $ s_2 $ from $ unvisited $
			\EndIf
			\State $ s_1 \leftarrow s_2 $
			\State $ s_2 \leftarrow $ Direct successor of $ s_1 $ (param: $ \textbf{x} $)
			\EndWhile
			\EndFor
			\If {\textit{unvisited} is empty}
			\State $ \textbf{S} \leftarrow \emptyset $
			\Else
			\State $ \textbf{S} \leftarrow $ all closed paths including tasks in $ unvisited $
			\EndIf
			\State \textbf{return} $ \textbf{S} $
			\EndProcedure
		\end{algorithmic}
		\label{alg:subtour}
	\end{algorithm}

	Whenever the callback is invoked by finding a new candidate incumbent solution that satisfies the constraints of the relaxed problem, the separation algorithm takes current decision variables denoted as $ \textbf{x}^*, \textbf{y}^* $, and $ \textbf{y}^{\textrm{NIN*}} $.
	Initially, the list variable \textit{unvisited} is defined and has an index of all tasks as an element.
	Then for each vehicle, the sample nodes are checked and are connected through the edges of a closed path passing through the depot and terminal clusters.
	The tasks of the connected sample nodes are removed from \textit{unvisited}, as well as the tasks for which their neighborhoods are indirectly visited.
	If \textit{unvisited} is an empty set after the above process, an empty set is assigned to the variable $ \textbf{S} $ and the procedure is terminated.
	Otherwise, all the closed paths which are passing through the tasks in \textit{unvisited} are found and saved in $ \textbf{S} $ to generate additional constraints after returning it because in the given solution there are some tasks that a fleet does not visit.

	\bibliography{bibtex_database}

\begin{thebibliography}{10}
\providecommand{\url}[1]{#1}
\csname url@samestyle\endcsname
\providecommand{\newblock}{\relax}
\providecommand{\bibinfo}[2]{#2}
\providecommand{\BIBentrySTDinterwordspacing}{\spaceskip=0pt\relax}
\providecommand{\BIBentryALTinterwordstretchfactor}{4}
\providecommand{\BIBentryALTinterwordspacing}{\spaceskip=\fontdimen2\font plus
\BIBentryALTinterwordstretchfactor\fontdimen3\font minus
  \fontdimen4\font\relax}
\providecommand{\BIBforeignlanguage}[2]{{%
\expandafter\ifx\csname l@#1\endcsname\relax
\typeout{** WARNING: IEEEtran.bst: No hyphenation pattern has been}%
\typeout{** loaded for the language `#1'. Using the pattern for}%
\typeout{** the default language instead.}%
\else
\language=\csname l@#1\endcsname
\fi
#2}}
\providecommand{\BIBdecl}{\relax}
\BIBdecl

\bibitem{ceccarelli2007micro}
N.~Ceccarelli, J.~J. Enright, E.~Frazzoli, S.~J. Rasmussen, and C.~J.
  Schumacher, ``Micro uav path planning for reconnaissance in wind,'' in
  \emph{American Control Conference, 2007. ACC'07}.\hskip 1em plus 0.5em minus
  0.4em\relax IEEE, 2007, pp. 5310--5315.

\bibitem{bortoff2000path}
S.~A. Bortoff, ``Path planning for uavs,'' in \emph{American Control
  Conference, 2000. Proceedings of the 2000}, vol.~1, no.~6.\hskip 1em plus
  0.5em minus 0.4em\relax IEEE, 2000, pp. 364--368.

\bibitem{shima2006multiple}
T.~Shima, S.~J. Rasmussen, A.~G. Sparks, and K.~M. Passino, ``Multiple task
  assignments for cooperating uninhabited aerial vehicles using genetic
  algorithms,'' \emph{Computers \& Operations Research}, vol.~33, no.~11, pp.
  3252--3269, 2006.

\bibitem{choi2010decentralized}
H.-L. Choi, A.~K. Whitten, and J.~P. How, ``Decentralized task allocation for
  heterogeneous teams with cooperation constraints,'' in \emph{American Control
  Conference (ACC), 2010}.\hskip 1em plus 0.5em minus 0.4em\relax IEEE, 2010,
  pp. 3057--3062.

\bibitem{ponda2012distributed}
S.~S. Ponda, L.~B. Johnson, A.~N. Kopeikin, H.-L. Choi, and J.~P. How,
  ``Distributed planning strategies to ensure network connectivity for dynamic
  heterogeneous teams,'' \emph{IEEE Journal on Selected Areas in
  Communications}, vol.~30, no.~5, pp. 861--869, 2012.

\bibitem{sundar2015exact}
K.~Sundar and S.~Rathinam, ``An exact algorithm for a heterogeneous, multiple
  depot, multiple traveling salesman problem,'' in \emph{Unmanned Aircraft
  Systems (ICUAS), 2015 International Conference on}.\hskip 1em plus 0.5em
  minus 0.4em\relax IEEE, 2015, pp. 366--371.

\bibitem{cai2014survey}
G.~Cai, J.~Dias, and L.~Seneviratne, ``A survey of small-scale unmanned aerial
  vehicles: Recent advances and future development trends,'' \emph{Unmanned
  Systems}, vol.~2, no.~02, pp. 175--199, 2014.

\bibitem{valavanis2014handbook}
K.~P. Valavanis and G.~J. Vachtsevanos, \emph{Handbook of unmanned aerial
  vehicles}.\hskip 1em plus 0.5em minus 0.4em\relax Springer Publishing
  Company, Incorporated, 2014.

\bibitem{obermeyer2012sampling}
K.~J. Obermeyer, P.~Oberlin, and S.~Darbha, ``Sampling-based path planning for
  a visual reconnaissance unmanned air vehicle,'' \emph{Journal of Guidance,
  Control, and Dynamics}, vol.~35, no.~2, pp. 619--631, 2012.

\bibitem{isaacs2013dubins}
J.~T. Isaacs and J.~P. Hespanha, ``Dubins traveling salesman problem with
  neighborhoods: a graph-based approach,'' \emph{Algorithms}, vol.~6, no.~1,
  pp. 84--99, 2013.

\bibitem{jang2016optimal}
C.~H.-J. Jang Dae-Sung and C.~Han-Lim, ``Optimal control-based uav path
  planning with dynamically-constrained tsp with neighborhoods,'' in
  \emph{International Conference on Control, Automation and Systems}, Jeju,
  Korea, 2017.

\bibitem{oberlin2010today}
P.~Oberlin, S.~Rathinam, and S.~Darbha, ``Today's traveling salesman problem,''
  \emph{IEEE robotics \& automation magazine}, vol.~17, no.~4, pp. 70--77,
  2010.

\bibitem{sundar2017path}
K.~Sundar, S.~Venkatachalam, and S.~G. Manyam, ``Path planning for multiple
  heterogeneous unmanned vehicles with uncertain service times,'' \emph{arXiv
  preprint arXiv:1702.07647}, 2017.

\bibitem{dantzig1954solution}
G.~Dantzig, R.~Fulkerson, and S.~Johnson, ``Solution of a large-scale
  traveling-salesman problem,'' \emph{Journal of the operations research
  society of America}, vol.~2, no.~4, pp. 393--410, 1954.

\bibitem{dubins1957curves}
L.~E. Dubins, ``On curves of minimal length with a constraint on average
  curvature, and with prescribed initial and terminal positions and tangents,''
  \emph{American Journal of mathematics}, vol.~79, no.~3, pp. 497--516, 1957.

\bibitem{le2012dubins}
J.~Le~Ny, E.~Feron, and E.~Frazzoli, ``On the dubins traveling salesman
  problem,'' \emph{IEEE Transactions on Automatic Control}, vol.~57, no.~1, pp.
  265--270, 2012.

\bibitem{oberlin2009transformation}
P.~Oberlin, S.~Rathinam, and S.~Darbha, ``A transformation for a heterogeneous,
  multiple depot, multiple traveling salesman problem,'' in \emph{American
  Control Conference, 2009. ACC'09.}\hskip 1em plus 0.5em minus 0.4em\relax
  IEEE, 2009, pp. 1292--1297.

\bibitem{zhang2014memetic}
X.~Zhang, J.~Chen, B.~Xin, and Z.~Peng, ``A memetic algorithm for path planning
  of curvature-constrained uavs performing surveillance of multiple ground
  targets,'' \emph{Chinese Journal of Aeronautics}, vol.~27, no.~3, pp.
  622--633, 2014.

\bibitem{cho2016informative}
D.-H. Cho, J.-S. Ha, S.~Lee, S.~Moon, and H.-L. Choi, ``Informative path
  planning and mapping with multiple uavs in wind fields,'' \emph{arXiv
  preprint arXiv:1610.01303}, 2016.

\bibitem{Kim2016cubature}
S.-H. Kim, L.~Negash, and H.-L. Choi, ``Cubature kalman filter based fault
  detection and isolation for formation control of multi-uavs,'' in \emph{IFAC
  Symposium on Intelligent Autonomous Vehicles}, Jun. 2016.

\bibitem{Choi2015informative}
\BIBentryALTinterwordspacing
H.-L. Choi and J.-S. Ha, ``Informative windowed forecasting of continuous-time
  linear systems for mutual information-based sensor planning,''
  \emph{Automatica}, vol.~57, pp. 97--104, Jul. 2015. [Online]. Available:
  \url{http://www.sciencedirect.com/science/article/pii/S0005109815001624}
\BIBentrySTDinterwordspacing

\bibitem{yuan2007optimal}
B.~Yuan, M.~Orlowska, and S.~Sadiq, ``On the optimal robot routing problem in
  wireless sensor networks,'' \emph{IEEE Transactions on Knowledge and Data
  Engineering}, vol.~19, no.~9, pp. 1252--1261, 2007.

\bibitem{liu2013path}
J.-S. Liu, S.-Y. Wu, and K.-M. Chiu, ``Path planning of a data mule in wireless
  sensor network using an improved implementation of clustering-based genetic
  algorithm,'' in \emph{Computational Intelligence in Control and Automation
  (CICA), 2013 IEEE Symposium on}.\hskip 1em plus 0.5em minus 0.4em\relax IEEE,
  2013, pp. 30--37.

\bibitem{gutin2006traveling}
G.~Gutin and A.~P. Punnen, \emph{The traveling salesman problem and its
  variations}.\hskip 1em plus 0.5em minus 0.4em\relax Springer Science \&
  Business Media, 2006, vol.~12.

\bibitem{bellmore1974transformation}
M.~Bellmore and S.~Hong, ``Transformation of multisalesman problem to the
  standard traveling salesman problem,'' \emph{Journal of the ACM (JACM)},
  vol.~21, no.~3, pp. 500--504, 1974.

\bibitem{laporte1980cutting}
G.~Laporte and Y.~Nobert, ``A cutting planes algorithm for the m-salesmen
  problem,'' \emph{Journal of the Operational Research society}, vol.~31,
  no.~11, pp. 1017--1023, 1980.

\bibitem{applegate2002solution}
D.~Applegate, W.~Cook, S.~Dash, and A.~Rohe, ``Solution of a min-max vehicle
  routing problem,'' \emph{INFORMS Journal on Computing}, vol.~14, no.~2, pp.
  132--143, 2002.

\bibitem{kivelevitch2013market}
E.~Kivelevitch, K.~Cohen, and M.~Kumar, ``A market-based solution to the
  multiple traveling salesmen problem,'' \emph{Journal of Intelligent \&
  Robotic Systems}, vol.~72, no.~1, p.~21, 2013.

\bibitem{baldacci2008routing}
R.~Baldacci, M.~Battarra, and D.~Vigo, ``Routing a heterogeneous fleet of
  vehicles,'' in \emph{The vehicle routing problem: latest advances and new
  challenges}.\hskip 1em plus 0.5em minus 0.4em\relax Springer, 2008, pp.
  3--27.

\bibitem{sundar2017algorithms}
K.~Sundar and S.~Rathinam, ``Algorithms for heterogeneous, multiple depot,
  multiple unmanned vehicle path planning problems,'' \emph{Journal of
  Intelligent \& Robotic Systems}, vol.~88, no. 2-4, pp. 513--526, 2017.

\bibitem{sundar2016generalized}
------, ``Generalized multiple depot traveling salesmen problem—polyhedral
  study and exact algorithm,'' \emph{Computers \& Operations Research},
  vol.~70, pp. 39--55, 2016.

\bibitem{salhi2014multi}
S.~Salhi, A.~Imran, and N.~A. Wassan, ``The multi-depot vehicle routing problem
  with heterogeneous vehicle fleet: Formulation and a variable neighborhood
  search implementation,'' \emph{Computers \& Operations Research}, vol.~52,
  pp. 315--325, 2014.

\bibitem{savla2005point}
K.~Savla, E.~Frazzoli, and F.~Bullo, ``On the point-to-point and traveling
  salesperson problems for dubins' vehicle,'' in \emph{American Control
  Conference, 2005. Proceedings of the 2005}.\hskip 1em plus 0.5em minus
  0.4em\relax IEEE, 2005, pp. 786--791.

\bibitem{ma2006receding}
X.~Ma and D.~A. Castanon, ``Receding horizon planning for dubins traveling
  salesman problems,'' in \emph{Decision and Control, 2006 45th IEEE Conference
  on}.\hskip 1em plus 0.5em minus 0.4em\relax IEEE, 2006, pp. 5453--5458.

\bibitem{cho2018heterogeneous}
D.-H. Cho, D.-S. Jang, and H.-L. Choi, ``Heterogeneous, multiple depot
  multi-uav path planning for remote sensing tasks,'' in \emph{AIAA SciTech
  Forum}, 01 2018.

\bibitem{helsgaun2000effective}
K.~Helsgaun, ``An effective implementation of the lin--kernighan traveling
  salesman heuristic,'' \emph{European Journal of Operational Research}, vol.
  126, no.~1, pp. 106--130, 2000.

\bibitem{obermeyer2009path}
K.~Obermeyer, ``Path planning for a uav performing reconnaissance of static
  ground targets in terrain,'' in \emph{AIAA Guidance, Navigation, and Control
  Conference}, 2009, p. 5888.

\bibitem{Zhang2014}
X.~Zhang, J.~Chen, B.~Xin, and Z.~Peng, ``A memetic algorithm for path planning
  of curvature-constrained uavs performing surveillance of multiple ground
  targets,'' \emph{Chinese Journal of Aeronautics}, vol.~27, no.~3, pp.
  622--633, 2014.

\bibitem{helsgaun2015solving}
K.~Helsgaun, ``Solving the equality generalized traveling salesman problem
  using the lin--kernighan--helsgaun algorithm,'' \emph{Mathematical
  Programming Computation}, vol.~7, no.~3, pp. 269--287, 2015.

\bibitem{freisleben1996genetic}
B.~Freisleben and P.~Merz, ``A genetic local search algorithm for solving
  symmetric and asymmetric traveling salesman problems,'' in \emph{Evolutionary
  Computation, 1996., Proceedings of IEEE International Conference on}.\hskip
  1em plus 0.5em minus 0.4em\relax IEEE, 1996, pp. 616--621.

\bibitem{renaud1998efficient}
J.~Renaud and F.~F. Boctor, ``An efficient composite heuristic for the
  symmetric generalized traveling salesman problem,'' \emph{European Journal of
  Operational Research}, vol. 108, no.~3, pp. 571--584, 1998.

\bibitem{snyder2006random}
L.~V. Snyder and M.~S. Daskin, ``A random-key genetic algorithm for the
  generalized traveling salesman problem,'' \emph{European journal of
  operational research}, vol. 174, no.~1, pp. 38--53, 2006.

\end{thebibliography}
	\bibliographystyle{IEEEtran}
	
\end{document}